\documentclass[a4paper,11pt]{article}

\usepackage{jheppub}

\numberwithin{equation}{section} \allowdisplaybreaks[4]

\newcommand{\bdm}{\begin{displaymath}}
\newcommand{\edm}{\end{displaymath}}
\newcommand{\beq}{\begin{equation}}
\newcommand{\eeq}{\end{equation}}
\newcommand{\bea}{\begin{eqnarray}}
\newcommand{\eea}{\end{eqnarray}}

\newcommand{\lt}{\left}
\newcommand{\rt}{\right}
\newcommand{\no}{\nonumber}
\newcommand{\nn}{\nonumber\\}
\newcommand{\ov}{\overline}
\newcommand{\eq}[1]{Eq.~(\ref{#1})}
\newcommand{\eqsand}[2]{Eqs.~(\ref{#1}) and (\ref{#2})}
\newcommand{\eqsto}[2]{Eqs.~(\ref{#1}) to (\ref{#2})}
\newcommand{\gev}{\mbox{GeV}}
\newcommand{\imag}{\mbox{Im\,}}
\newcommand{\real}{\mbox{Re\,}}

\newcommand{\Bbar}{\bar{B}}

\newcommand{\bbs}{\ensuremath{B_s\!-\!\Bbar{}_s\,}}

\newcommand{\bra}[1]{\ensuremath{\langle #1 |}}
\newcommand{\ket}[1]{\ensuremath{| #1 \rangle }}
\newcommand{\fig}[1]{Fig.~\ref{#1}}
\newcommand{\tab}[1]{Tab.~\ref{#1}}
\newcommand{\lqcd}{\Lambda_{\textit{\scriptsize{QCD}}}}

\newcommand{\dm}{\ensuremath{\Delta M}}
\newcommand{\dg}{\ensuremath{\Delta \Gamma}}
\newcommand{\epm}[2]{
 \raisebox{-0.5ex}{\shortstack[l]{$\scriptstyle+#1$\\$\scriptstyle-#2$}}}

\def \be{\begin{equation}}
\def \ee{\end{equation}}

\catcode`@=11
\def\marginnote#1{}
\newcount\hour
\newcount\minute
\newtoks\amorpm
\hour=\time\divide\hour by60 \minute=\time{\multiply\hour by60
\global\advance\minute by-\hour}
\edef\standardtime{{\ifnum\hour<12 \global\amorpm={am}%
        \else\global\amorpm={pm}\advance\hour by-12 \fi
        \ifnum\hour=0 \hour=12 \fi
        \number\hour:\ifnum\minute<10 0\fi\number\minute\the\amorpm}}
\edef\militarytime{\number\hour:\ifnum\minute<10 0\fi\number\minute}
\def\draftlabel#1{{\@bsphack\if@filesw {\let\thepage\relax
   \xdef\@gtempa{\write\@auxout{\string
      \newlabel{#1}{{\@currentlabel}{\thepage}}}}}\@gtempa
   \if@nobreak \ifvmode\nobreak\fi\fi\fi\@esphack}
        \gdef\@eqnlabel{#1}}
\def\@eqnlabel{}
\def\@vacuum{}
\def\draftmarginnote#1{\marginpar{\raggedright\scriptsize\tt#1}}
\def\draft{\oddsidemargin 0.0truein
        \def\@oddfoot{\sl preliminary draft \hfil
        \rm\thepage\hfil\sl\today\quad\militarytime}
        \let\@evenfoot\@oddfoot \overfullrule 3pt
        \let\label=\draftlabel
        \let\marginnote=\draftmarginnote
   \def\@eqnnum{(\theequation)\rlap{\kern\marginparsep\tt\@eqnlabel}%
\global\let\@eqnlabel\@vacuum}  } \catcode`@=12

\title{\boldmath Towards next-to-next-to-leading-log
    accuracy for the width difference in the \bbs\ system:\\[2mm] fermionic
    contributions to order $(m_c/m_b)^0$ and $(m_c/m_b)^1$}

\author[a]{H.M. Asatrian,}
\author[a]{A. Hovhannisyan,}
\author[b]{U. Nierste}
\author[a]{A. Yeghiazaryan}

\affiliation[a]{Yerevan Physics Institute, 0036 Yerevan, Armenia}
\affiliation[b]{Institut f\"ur Theoretische Teilchenphysik, Karlsruhe
  Institute of Technology, Engesserstra\ss e 7, 76128 Karlsruhe,
  Germany}

\emailAdd{hrachia@yerphi.am}
\emailAdd{artyom@yerphi.am}
\emailAdd{ulrich.nierste@kit.edu}
\emailAdd{arseny@yerphi.am}

\abstract{We calculate a class of three-loop Feynman diagrams which contribute to
the next-to-next-to-leading logarithmic approximation for the width
difference $\Delta\Gamma_{s}$ in the $B_s-\bar{B}_s$ system.  The
considered diagrams contain a closed fermion loop in a gluon propagator
and constitute the order $\alpha_s^2 N_f$, where $N_f$ is the number of
light quarks.  Our results entail a considerable correction in that
order, if $\Delta\Gamma_{s}$ is expressed in terms of the pole mass of
the bottom quark. If the $\overline{\rm MS}$ scheme is used instead, the
correction is much smaller. As a result, we find a decrease of the
scheme dependence. Our result also indicates that the usually quoted
value of the NLO renormalization scale dependence underestimates the
perturbative error.}

\arxivnumber{1709.02160}
\begin{document}
\maketitle
\flushbottom

\section{Introduction}
\bbs\ oscillations are governed by the $2\times 2$ matrix
$M-i\Gamma/2$, which contains the mass matrix $M=M^\dagger$ and the
decay matrix $\Gamma=\Gamma^\dagger$. By diagonalising $M-i\Gamma/2$
one finds the mass eigenstates $B_L$ and $B_H$ with the subscripts
denoting ``light'' and ``heavy'', respectively. The eigenvalues
$M_L-i \Gamma_L/2$ and $M_H-i \Gamma_H/2$ define masses and decay
width of $B_L$ and $B_H$. The time-dependent states $B_L(t)$ and
$B_H(t) $ each obey exponential decay laws with decay constants
$\Gamma_L$ and $\Gamma_H$.  By transforming back to the flavour
basis $(B_s, \Bbar_s)$ one finds the familiar damped oscillations
between these flavour eigenstates.  The mixing problem involves five
observables:
\begin{align}
M &= \frac{M_L+M_H}{2},  &
\Gamma &= \frac{\Gamma_L+\Gamma_H}{2}, & 
\dm &= M_H-M_L, & \dg &= \Gamma_L-\Gamma_H, &
\end{align}
and the CP asymmetry in flavour-specific decays, $a_{\rm fs}$, which
quantifies CP violation in mixing. The mass difference $\dm =
(17.757 \pm 0.021)\, \mbox{ps}^{-1}$ \cite{hfag} has been determined
very precisely by the CDF \cite{Abulencia:2006ze} and LHCb
\cite{Aaij:2013mpa} experiments from the \bbs\ oscillation
frequency. The experimental value of the width difference
\cite{hfag},
\begin{align}
\dg^{\rm exp}  &= (0.089 \pm 0.006)\, \mbox{ps}^{-1}
\label{eq:dgexp}
\end{align}
is an average of measurements by LHCb \cite{lhcb}, \cite{lhcb2}, ATLAS
\cite{Aad:2016tdj}, CMS \cite{Khachatryan:2015nza}, and
CDF \cite{Aaltonen:2012ie}. The average mass
$M\equiv M_{B_s}$ and the average width $\Gamma$ of the mass
eigenstates are simply given by the diagonal elements of $M$ and
$\Gamma$ as $M=M_{11}=M_{22}$ and $\Gamma=\Gamma_{11}=\Gamma_{22}$.
The remaining physical quantities in $M-i \Gamma/2$ are $|M_{12}|$,
$|\Gamma_{12}|$, and the CP-violating phase $\phi_{12} = \arg
(-M_{12}/\Gamma_{12})$. These are related to $\dm$, $\dg$, and
$a_{\rm
  fs}$ as
\begin{align}
\dm &=\, 2|M_{12}|, & \dg &=\, 2|\Gamma_{12}| \cos \phi_{12} = \, -
\dm \,\real \frac{\Gamma_{12}}{M_{12}}, \nn && a_{\rm fs} &=\,
\frac{|\Gamma_{12}|}{|M_{12}|} \sin  \phi_{12}  = \, \imag
\frac{\Gamma_{12}}{M_{12}} .
\end{align}
In these formulas $|\Gamma_{12}| \ll |M_{12}|$ and $|\dg|\ll |\dm|$
is used. Within the Standard Model (SM) one finds
$\phi_{12}=0.24^\circ\pm 0.06 ^\circ$
\cite{Beneke:2003az,NiersteNLONB,Lenz:2011ti,Nierste:2012qp}, which
permits to set $\cos \phi_{12} = 1$ in the SM prediction for $\dg$.


For the calculation of $\Gamma_{12}$ one employs an operator product
expansion, the heavy quark expansion (HQE) \cite{hqe}-\cite{hqe4}, which results in
a systematic expansion of $\Gamma_{12}$ in powers of $\lqcd/m_b\sim 0.1$
and $\alpha_s(m_b)\sim 0.2$. $\Gamma_{12}$ has been calculated to
next-to-leading order (NLO) in both $\lqcd/m_b$ \cite{HNSBsBsbar} and
$\alpha_s(m_b)$
\cite{NiersteNLO,Ciuchini:2003ww,Beneke:2003az,NiersteNLONB}. The
leading-power (i.e.~$(\lqcd/m_b)^0$) term involves two $|\Delta B|=2$
operators ($B$ denotes the beauty quantum number)
\begin{eqnarray}
  Q=(\bar{s}_ib_i)_{V-A}\;(\bar{s}_jb_j)_{V-A},
 \qquad\qquad
 \tilde{Q}_S=(\bar{s}_ib_j)_{S-P}\;(\bar{s}_jb_i)_{S-P}. \label{eq:defops}
\end{eqnarray}
Here the $i,j$ are colour indices and $V\pm A$ means
$\gamma_{\mu}(1\pm \gamma_5)$ while $S\pm P$ stands for $(1\pm
\gamma_5)$. The hadronic matrix elements, which must be calculated
with non-perturbative methods, are usually parameterized as
\begin{eqnarray}
  \bra{B_s} Q (\mu_2) \ket{\ov B_s}  &=&
   \frac{8}{3} M^2_{B_s}\, f^2_{B_s} B(\mu_2)
  \qquad
  \bra{B_s} \widetilde Q_S (\mu_2)\ket{\ov B_s} \;=\; \frac{1}{3}  M^2_{B_s}\,
  f^2_{B_s} \widetilde B_S^\prime (\mu_2).
      \label{eq:defb}
\end{eqnarray}
Here $f_{B_s}$ is the $B_s$ decay constant and $\mu_2={\cal O}(m_b)$
is the renormalization scale at which the matrix elements are
calculated. In a lattice-gauge theory calculation $\mu_2$ is the
scale at which the lattice-continuum matching is performed. In the
expression for $\Gamma_{12}$ the matrix elements of \eq{eq:defb} are
multiplied by perturbative Wilson coefficients which also depend on
$\mu_2$ such that the dependence on the unphysical scale $\mu_2$
cancels from $\Gamma_{12}$. In the same way the dependence on the
renormalization scheme cancels between the  Wilson coefficients and
$B(\mu_2)$, $\widetilde B_S^\prime (\mu_2)$. In this paper we use the scheme of
Ref.~\cite{NiersteNLO}.

$\dg$ is proportional to $m_b^2$ and the theoretical prediction
depends on the renormalization scheme chosen for $m_b$ (for a
detailed discussion see Ref.~\cite{NiersteNLONB}) and further on the
scale $\mu_1={\cal O}(m_b)$ at which the $|\Delta B|=1$ Wilson
coefficients are evaluated. Both dependences are unphysical and
diminish order-by-order in perturbation theory. At NLO the scheme
and scale dependence is still sizable and indicates that higher
orders of $\alpha_s$ should be calculated.  With up-to-date values
for quark masses and the elements of the Cabibbo-Kobayashi-Maskawa
(CKM) matrix (stated below in Sec.~\ref{sec:num}) one finds
\begin{align} 
\dg &=\; (1.74 \pm 0.24) \,
          f_{B_s}^2 B  \; +\;  (0.40\pm 0.05) f_{B_s}^2 \widetilde
          B_S^\prime
         \; +\;   (-0.65 \pm 0.35) \, f_{B_s}^2 \label{eq:dgp}
\end{align}
in the scheme using the pole mass definition of $m_b$ in the
prefactor of $\dg$. Here and in the following the hadronic
parameters are understood at $\mu_2=m_b$. The last term in
\eq{eq:dgp} is the $\lqcd/m_b$ correction. If instead the $\ov{\rm
MS}$ scheme is used for $m_b$ one finds
\begin{align} 
\dg &=\;  (1.86\pm 0.08) \,
          f_{B_s}^2 B  \; +\;  (0.42\pm 0.01) f_{B_s}^2 \widetilde B_S^\prime
         \; +\; (-0.55 \pm 0.29) \, f_{B_s}^2. \label{eq:dgb}
\end{align}
The errors quoted in the brackets in \eqsand{eq:dgp}{eq:dgb} are
found by varying $\mu_1$ between $m_b/2$ and $2 m_b$.
Ref.~\cite{Lenz:2011ti} has quoted all results for the scheme of
\eq{eq:dgb}, while in Ref.~\cite{Nierste:2012qp} the average of
results in the two schemes has been given.  A recent lattice
calculation \cite{Bazavov:2016nty} has found
\begin{align}
  f_{B_s}^2 B &= \, [0.224 \, \gev]^2   \,
    (1.00 \pm 0.06), & \qquad
 f_{B_s}^2 \widetilde B_S^\prime &= \, [0.224 \, \gev] ^2\,
  (1.83 \pm 0.19) \label{eq:latt1}
\end{align}
Here we have added two errors from different sources in quadrature.
Ref.~\cite{Bazavov:2016nty} has also calculated some of the matrix
elements appearing at order $\lqcd/m_b$ and these results went into
the last terms of \eqsand{eq:dgp}{eq:dgb}. With
$f_{B_s}=0.224\,\gev$ and neglecting the correlation of the
uncertainties in $B$ and $\widetilde B_S^\prime$ we find
\begin{align}
\dg &= \lt( 0.0913 \pm 0.020_{\rm scale} \pm 0.006_{B,\widetilde
B_S}
       \pm  0.017_{\lqcd/m_b} \rt) \, \gev
\qquad \mbox{(pole)} \nn
\dg &= \lt( 0.104 \pm 0.008_{\rm scale} \pm
   0.007_{B,\widetilde B_S}
   \pm 0.015_{\lqcd/m_b} \rt)\, \gev
\qquad \mbox{($\ov{\rm MS}$)} \label{eq:numintro}
\end{align}
From \eq{eq:numintro} we observe that the both scale and scheme
dependences exceed the uncertainties from the hadronic parameters
$B$ and $\widetilde B_S$. Furthermore, the theoretical uncertainty
inferred from these dependences is larger than the present
experimental error. This calls for a NNLO calculation of the
perturbative coefficients multiplying $Q$ and $\widetilde Q_S$.  In
this paper we present the first step in this direction, the
calculation of the terms of order $\alpha_s^2 N_f$, where $N_f$ is
the number of quark flavours, neglecting quadratic and higher powers
of $m_c/m_b$. \eq{eq:numintro} will further improve from a future
calculation of the NLO corrections to the $\lqcd/m_b$ part and
progress in the lattice calculations of the hadronic matrix elements
appearing in this order. The contributions of order $(\lqcd/m_b)^2$,
however, have been estimated to be small
\cite{NiersteNLONB,Badin:2007bv}. The theoretical prediction can be
further refined, if $\dg$ is predicted from the ratio $\dg/\dm$ and
the experimental value of $\dm$, which is proportional to $f_{B_s}^2
B$. This procedure eliminates the uncertainty associated with $B$
altogether, at the price of making the prediction sensitive to
possible new physics in $\dm$. From \eqsand{eq:dgp}{eq:dgb} one
realises that the numerically dominant term in $\dg/\dm$ will not
contain any hadronic parameter \cite{NiersteNLONB}. This feature
also alleviates the problem that the lattice-continuum matching is
currently only known to NLO. 

This paper is organized as follows: In the following section we
summarize the theoretical framework of the calculation. In
Sec.~\ref{sec:rir} we describe details of the renormalization
procedure and the regularization of infrared singularities.
We present our analytical results in Sec.~\ref{sec:cf} and
perform a phenomenological analysis in Sec.~\ref{sec:num}.
Finally we conclude. Results for matrix elements and
master integrals needed for the calculation are relegated to
the appendix.

\section{\bf Theoretical framework\label{sec:tf}}
The effective $\Delta B=1$ weak Hamiltonian, relevant for our
calculation, is the following \cite{Buchalla:1995vs}
\begin{eqnarray}
\label{Heff} H^{\Delta
B=1}_{eff}=\frac{G_F}{\sqrt{2}}V^*_{cs}V_{cb}\left\{
\sum^6_{i=1}C_iO_i + C_{8}O_{8}\right\} \; +\; \mbox{H.c.},
\label{eq:heff}
\end{eqnarray}
with the operators
\begin{eqnarray}\label{OperBasis}
\nonumber O_1=(\bar{s}_ic_j)_{V-A}\;(\bar{c}_jb_i)_{V-A}, \qquad\qquad
O_2=(\bar{s}_ic_i)_{V-A}\;(\bar{c}_jb_j)_{V-A},
\\
\nonumber O_3=(\bar{s}_ib_i)_{V-A}\;(\bar{q}_jq_j)_{V-A}, \qquad\qquad
O_4=(\bar{s}_ib_j)_{V-A}\;(\bar{q}_jq_i)_{V-A},
\\
O_5=(\bar{s}_ib_i)_{V-A}\;(\bar{q}_jq_j)_{V+A}, \qquad\qquad
O_6=(\bar{s}_ib_j)_{V-A}\;(\bar{q}_jq_i)_{V+A},
\\
\nonumber
O_{8}=\frac{g_s}{8\pi^2}m_b\bar{s}_i\sigma^{\mu\nu}(1-\gamma_5)T_{ij}^a
b_jG^a_{\mu\nu}.~~~~~~~~~~~~~~~~~
\end{eqnarray}
Here the $i,j$ are colour indices and summation over $q = u, d, s,
c, b$ is implied. $V\pm A$ refers to $\gamma_{\mu}(1\pm \gamma_5)$
and $S\pm P$ (which we need below) to $(1\pm \gamma_5)$.
$C_1,\ldots,C_6$ and $C_8$ are the corresponding Wilson coefficient
functions. $G_F$ is the Fermi constant and $V_{jk}$ denotes an element
of the CKM matrix.  Cabbibo-suppressed contributions proportional to
$V^*_{ub}V_{us}$ are neglected in (\ref{Heff}).

To find $\dg \simeq 2|\Gamma_{12}|$ we must calculate
\begin{eqnarray}
\Gamma_{12} &=& \mbox{Abs}
 \bra{B_s} \,i\!\int d^4x\ T\,{\cal H}_{eff}(x){\cal H}_{eff}(0)
 \ket{\Bbar_s}, \label{eq:full}
\end{eqnarray}
where `Abs' denotes the absorptive part of the matrix element
and $T$ denotes time ordering.
The HQE expresses \eq{eq:full} in terms of matrix elements
of local operators. The leading term (in powers of $\lqcd/m_b$) reads
\begin{eqnarray}
\Gamma_{12} &=& \frac{G_F^2 m_b^2}{24 \pi\, M_{B_s}} \lt(V_{cb}^*V_{cs}\rt)^2
    \lt[ \, G \, \bra{B_s } Q \ket{\Bbar_s} \; - \;
          G_S \,\bra{B_s} Q_S \ket{\Bbar_s}
    \rt]  
\label{defg}
\end{eqnarray}
Using the notation of Refs.~\cite{NiersteNLO,Beneke:2003az,
  NiersteNLONB}, the coefficients $G$ and $G_S$ are further
decomposed as
\begin{eqnarray}
G \;=\; F + P, \qquad
G_S \; = \; - F_S -  P_S
    . \label{deffp}
\end{eqnarray}
Here $F$ and $F_S$ are the contributions from the current-current
operators $Q_{1,2}$ while the small coefficients $P$ and $P_S$ stem
from the penguin operators $Q_{3-6}$ and $Q_8$. The coefficients
$G,G_S$ are calculated by expressing the bilocal matrix
elements
\begin{eqnarray}
  \mbox{Abs}\,
 \langle \;i\!\int d^4x\ T\, Q_i (x) Q_j(0) \, \rangle, \label{eq:fab}
\end{eqnarray}
(``full theory'') in terms of the local matrix elements $\langle Q
\rangle$, $\langle Q_S \rangle$ (``effective theory''), the coefficients of
the latter are the desired coefficients. Since $G,G_S$ are
short-distance quantities, this matching calculation can be done
order-by-order in perturbation theory, with quarks instead of mesons as
\begin{figure}[t]
\vspace{-1.8cm} 
\includegraphics[width=0.85\textwidth]{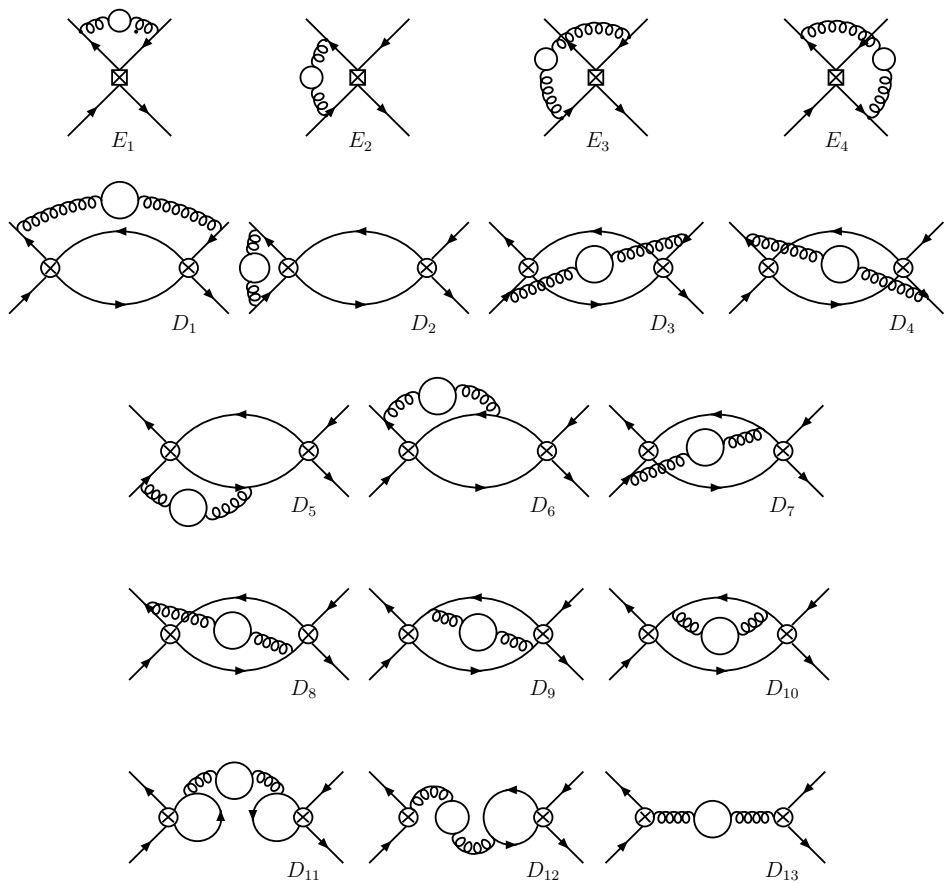}
 \vspace{-6cm}
 \caption{Diagrams $D_1 - D_{13}$ constitute the
   ${O}(\alpha_s^2 N_f)$ corrections to \eq{eq:fab}.  $E_1-E_4$ are the
   corresponding corrections to the matrix elements of local $\Delta
   B=2$ operators, which are required for a proper factorization of
   infrared divergences.  Not displayed are $E_1'$, $E_2'$,
   $D_1'$, $D_2'$, $D_5'$, $D_6'$, $D_7'$, $D_8'$, $D_{10}'$ and
   $D_{12}'$ which are obtained by rotating the corresponding diagrams
   by $180^0$ and diagrams associated with QCD penguin operators.  The closed
   fermion loop contains massive $c$, $b$ quarks and massless $u$, $d$,
   $s$ quarks. The charm loop involves terms of order $m_c/m_b$, so that
   the charm mass cannot be neglected here. However, in the charm quark
   lines attached to a weak vertex we set the charm quark mass to zero,
   which induces an error of order $m_c^2/m_b^2$. \label{fig1}}
~\\[-5mm]\hrule
\end{figure}
external states in \eq{eq:fab}.  The NLO result of
Refs.~\cite{NiersteNLO,Ciuchini:2003ww,Beneke:2003az,NiersteNLONB}
involves \eq{eq:fab} at the two-loop level for $i,j=1,2$.  The
chromomagnetic operator $O_8$ is proportional to the strong coupling
$g_s$, so that for $i=8$ or $j=8$ a one-loop calculation is sufficient
for NLO accuracy. It is further customary to count the small penguin
Wilson coefficients $C_{3-6}$ as ${\cal O}(\alpha_s)$ and only one-loop
diagrams are considered for $i\geq 3$ or $j\geq 3$.

The first ingredient of an NNLO result are the Wilson coefficients of
the $\Delta B=1$ weak Hamiltonian in \eq{eq:heff}.  The NNLO Wilson
coefficients involve the three-loop anomalous dimension matrix governing
the renormalization-group evolution of $C_{1-6,8}$ from the electroweak
scale down to the scale $\mu_1\sim m_b$, at which the matrix elements in
\eq{eq:fab} are evaluated. The NNLO effective hamiltonian has been
calculated in Refs.~\cite{Gorbahn:2004my,Gorbahn:2005sa}, albeit in a
different operator basis than the one in \eq{OperBasis}, which is used
in the NLO calculations of
Refs.~\cite{NiersteNLO,Ciuchini:2003ww,Beneke:2003az,NiersteNLONB} and in
this paper.

The NNLO contributions presented in this paper all involve a closed
quark loop and would be dominant in the case of a large number $N_f$ of
light quarks. However, the limit $N_f\to \infty$ is in conflict with
asymptotic freedom of QCD, as the first term $\beta_0$ of the QCD
$\beta$ function would change sign. It has been suggested to trade $N_f$
for $\beta_0$, so that the $\alpha_s^2 N_f$ term is replaced by a term
of order $\alpha_s^2 \beta_0$ (naive non-abelianization
\cite{Brodsky:1982gc,Beneke:1994qe}). In some applications this
procedure gives a good approximation to the full $\alpha_s^2$ term.
However, in quantities involving effective four-quark operators, it is
pure speculation whether the original $\alpha_s^2 N_f$ term or its
naively non-abelianized version $\propto \alpha_s^2 \beta_0$
approximates the full result in a better way, because neither term
cancels the scheme dependence of the operator renormalization. That is,
in one scheme the $\alpha_s^2 N_f$ term may be a good approximant, while
in another one the $\alpha_s^2 \beta_0$ term does better, or neither of
them is sensible. For the standard NDR renormalization scheme used by
us, e.g.\ the calculation in Ref.~\cite{Asatrian:2010rq} revealed that
the $\alpha_s^2 \beta_0$ term is not a good approximation to the full
result. In light of this finding we do not advocate the use of naive
non-abelianization in our case.  Nonetheless, the $\alpha_s^2 N_f$
portion of the full NNLO result is gauge invariant and therefore a
meaningful quantity. One can also overcome the scheme-dependence issue
by only keeping the $\alpha_s^2 N_f$ terms of the NNLO correction to the
RG-improved Wilson coefficients. However, we find that applying this
procedure to the known NLO result gives a poor approximation, so that we
refrain from using it.

The desired $\alpha_s^2 N_f $ contribution requires the calculation of
the diagrams in \fig{fig1}. We formally distinguish the charm mass
in the lines attached to an effective operator (i.e.\ to a weak vertex)
from that in the charm loop correcting
the gluon propagator: The latter give rise to corrections which are
linear in $m_c/m_b$ and we keep a non-zero charm mass in these loops.
On the contrary, the dependence on the charm mass arising from the lines
in which the charm originates from a weak vertex is only quadratic and
we use $m_c=0$ for these lines.
Denoting the $\ov{\rm MS}$-renormalized mass of the quark $q$ with
$m_q(\mu)$, where $\mu$ is the renormalization scale, we define
\begin{align}
z=&\, \frac{m_c^2(m_c)}{m_b^2(m_b)} = 0.095,\qquad \mbox{and}\qquad
\bar z= \, \frac{m_c^2(m_b)}{m_b^2(m_b)} = 0.048.
\label{eq:defz}
\end{align}
If the LO and NLO terms are expressed in terms of $z$, the error
associated with the above approximation is of order $\alpha_s^2 N_f z
\log^2 z$. If, however, one uses $\bar z$ instead, the approximation
only inflicts an error of order $\alpha_s^2 N_f \bar z $ and the
logarithmic terms $\alpha_s^n z \log^n z$, $z=1,\ldots$ are summed to all
orders. This feature has been studied in
Ref.~\cite{Beneke:2002rj,NiersteNLONB}.  The NLO result for $\dg$
expressed through $\bar z$ is numerically very well reproduced if $\bar
z$ is set to zero in the NLO correction. Since we discard terms of order
$\alpha_s^2 \bar z$, one may also expand the $z$-dependence from the
charm quark loop to order $z\log z$ and neglect terms of order $z$ and
higher. We calculate the tree-loop diagrams with charm loop indeed
as an expansion in $z$, but keep all terms to order $z^3$, to check
whether the expansion is numerically under good control. Furthermore, a
future NNLO calculation keeping higher powers of $z$ terms will benefit
from these results.

\section{\bf Renormalization and infrared regularization\label{sec:rir}}
In this section we specify our renormalization scheme,
present the various counterterms, and clarify the
regularization procedure used to isolate infrared (IR) divergences.
The latter factorize between the full-theory and effective-theory
diagrams (see \fig{fig1}) and render the desired Wilson
coefficients IR-finite.

For the three-loop diagrams involving two insertions of $O_{1,2}$ we
  need $C_{1,2}$ at NNLO (i.e.\ calculated with three-loop anomalous
  dimensions). The result of Ref.~\cite{Gorbahn:2004my} has been
  transformed to the traditional operator basis in \eq{OperBasis} in
  Ref.~\cite{Buras:2006gb} and we use the result of this paper.
We renormalize the operators in the usual naive dimensional regularization
(NDR) scheme. To fully specify the scheme one must further
define the evanescent operators \cite{NiersteEvan}. In
Ref.~\cite{Buras:2006gb} the usual NLO definition of these operators has
been extended to NNLO in such a way that the diagonal
RG evolution of $O_2\pm O_1$ is maintained at NNLO. For our calculation
we must specify the evanescent operators related to
$Q$ and $\tilde{Q}_S$: Their operational definition involves the
following replacements in
the $D$-dimensional Dirac-structures
($D=4-2\epsilon$):
\begin{eqnarray}
\label{2gamma} && [\gamma^{\mu}\gamma^{\nu}(1-\gamma_5)]_{ij}
[\gamma_{\nu}\gamma_{\mu}(1-\gamma_5)]_{kl} \quad\to\quad (8-8\epsilon)
[1-\gamma_5]_{il} [1-\gamma_5]_{kj} + 4\epsilon^2 [1-\gamma_5]_{ij}
[1-\gamma_5]_{kl},~~~
\\[1mm]
\label{3gamma} &&
[\gamma^{\mu}\gamma^{\alpha}\gamma^{\nu}(1-\gamma_5)]_{ij}
[\gamma_{\nu}\gamma_{\alpha}\gamma_{\mu}(1-\gamma_5)]_{kl}
\quad\to\quad
(4-8\epsilon+4\epsilon^2) [\gamma^{\mu}(1-\gamma_5)]_{ij}
[\gamma_{\mu}(1-\gamma_5)]_{kl}.
\end{eqnarray}
These relations, as well as their colour-flipped counterparts, extend
the result of Ref.~\cite{NiersteNLO} to order $\epsilon^2$. Formally,
the evanescent operator $E_1[Q]$ (see Ref.~\cite{NiersteEvan}) is
defined as the difference between the expression on the left and on the
right of the arrow in \eq{3gamma}, supplemented with the quark field
operators on the left and right of the Dirac structures, and analogously
\eq{2gamma} defines $E_1[\tilde{Q}_S]$. At NNLO the $\epsilon^2$ terms
matter, and these are chosen to preserve the Fierz symmetry, i.e. the
two-loop matrix elements of $Q$ and $\tilde{Q}_S$ are equal to the
matrix elements of the operators obtained from $Q$ and $\tilde{Q}_S$ by
4-dimensional Fierz transformations.

In a first step of the calculation the diagrams contributing to
\eq{eq:fab} generate three effective operators, $Q$, $\tilde{Q}_S$
and  $Q_S= (\bar{s}_ib_i)_{S-P}\;(\bar{s}_jb_j)_{S-P}$.
However, one linear combination of $Q$, $Q_S$, and $\tilde{Q}_S$ is $1/m_b$
suppressed \cite{HNSBsBsbar}, so that one can choose any two
of them in the leading-power result addressed in this paper.
The $1/m_b$-suppressed operator reads
\begin{eqnarray}
\label{R0} R_0 \equiv Q_S+\alpha_1 \tilde{Q}_S+\alpha_2\frac{1}{2}Q,
\end{eqnarray}
with $\alpha_{1,2}=1$ at LO. In Ref.~\cite{NiersteNLO} it was found that
 $\alpha_{1,2}$ receive corrections of order $\alpha_s$. To our order
$\alpha_s^2 N_f$ and in the scheme defined by \eqsand{2gamma}{3gamma}
these coefficients read:
\begin{eqnarray}
&& \nonumber \alpha_1 = 1+\frac{\alpha_s(\mu_2)}{4\pi}C_f
\left(12\log\frac{\mu_2}{m_b}+6\right)
+\frac{\alpha_s^2(\mu_2)}{(4\pi)^2}C_f \left[N_H \left(-\frac{52}{3}
\log \frac{\mu_2}{m_b}-8 \log
^2\frac{\mu_2}{m_b}-\frac{427}{18}+\frac{8 \pi ^2}{3}\right)\right.
\\
&& \nonumber \hspace{0.5cm} +N_V \left(-\frac{52}{3}
\log\frac{\mu_2}{m_b}-8 \log
^2\frac{\mu_2}{m_b}-\frac{211}{18}-\frac{4 \pi^2}{3}+4 \pi ^2
\sqrt{z}-24 z+4 \pi ^2 z^{3/2}\right.
\\
&& \left. \nonumber \hspace{1.7cm} -\frac{1}{9} z^2 \left(151+12 \pi
^2-78 \log z+18 \log ^2 z\right) +\frac{8}{75} z^3 (19-10 \log
z)\right)
\\
&& \left. \hspace{0.5cm} +N_L \left(-\frac{52}{3} \log
\frac{\mu_2}{m_b}-8 \log ^2 \frac{\mu_2}{m_b}-\frac{211}{18}-\frac{4
\pi^2}{3}\right) \right], \label{eq:al1}
\end{eqnarray}
\begin{eqnarray}
&& \nonumber \alpha_2 = 1+\frac{\alpha_s(\mu_2)}{4\pi}C_f
\left(6\log\frac{\mu_2}{m_b}+\frac{13}{2}\right)
+\frac{\alpha_s^2(\mu_2)}{(4\pi)^2}C_f \left[N_H \left(-\frac{26}{3}
\log \frac{\mu_2}{m_b}-4 \log
^2\frac{\mu_2}{m_b}-\frac{217}{18}+\frac{4 \pi ^2}{3}\right) \right.
\\
&& \nonumber \hspace{0.5cm} +N_V \left(-\frac{26}{3}
\log\frac{\mu_2}{m_b}-4 \log
^2\frac{\mu_2}{m_b}-\frac{109}{18}-\frac{2 \pi^2}{3}+2 \pi ^2
\sqrt{z}-12 z+2 \pi ^2 z^{3/2}\right.
\\
&& \left. \nonumber \hspace{1.7cm} -\frac{1}{18}z^2 \left(18 \log ^2
z-78 \log z+12 \pi ^2+151\right)+\frac{4}{75} z^3 (19-10 \log
z)\right)
\\
&& \left. \hspace{0.5cm} +N_L \left(-\frac{26}{3} \log
\frac{\mu_2}{m_b}-4 \log ^2\frac{\mu_2}{m_b}-\frac{109}{18}-\frac{2
\pi^2}{3}\right) \right]. \label{eq:al2}
\end{eqnarray}
Here $C_f =4/3$ is a colour factor and $\mu_2$ is the scale at which the
operators in Eq.(\ref{R0}) are defined.  $N_H=1$, $N_V=1$, and $N_L=3$
count the numbers of $b$, $c$, and light ($u,d,s$) quarks,
respectively. The redundant parameters $N_{H,V}$ are introduced for an
easier recognition of the various contributions in the formulae for the
coefficients.  The results for $\alpha_{1,2}$ are further expanded in
$z$ to the third order.  Later we will have to express $\alpha_s(\mu_2)$
in terms of $\alpha_s(\mu_1)$, which occurs in the Wilson coefficients.
To this end one can use the following formula:
\begin{eqnarray}
\label{scale}
\alpha_s(\mu_2)=\alpha_s(\mu_1)+\frac{\alpha_s^2(\mu_1)}{2 \pi} \beta_0 \log
\frac{\mu_1}{\mu_2}.
\end{eqnarray}

One may freely choose two of the three operators $Q$, $Q_S$, and
$\tilde{Q}_S$. The choice of the basis $Q,\tilde{Q}_S$ leads to
numerically more stable results \cite{NiersteNLONB} than the choice
$Q,Q_S$ and renders the unknown NLO corrections proportional to $\langle
R_0\rangle $ color-suppressed.  Nevertheless the NNLO calculation is
more convenient in the latter basis and one may easily transform the
result between the bases by using \eqsto{R0}{eq:al2}.

We next discuss the infrared regularization.
For the gluon propagator we use the following expression (similar to
the W boson propagator in an $R_\xi$ gauge with $\xi=0$)
\begin{eqnarray}
&& \frac{-i
\delta_{ab}}{k^2-m_g^2+i\epsilon}\left(g_{\mu\nu}-\frac{k_{\mu}
k_{\nu}}{k^2}\right),
\end{eqnarray}
where $m_g$ is a gluon mass.
Our choice of a gluon mass as IR regulator instead of
using dimensional regularization has two advantages:
In the matching procedure we do not need the $\epsilon$ and
$\epsilon^2$ parts of NLO and LO Wilson coefficients and
the disapperance of $m_g$ from the Wilson coefficients provides a
non-trivial check of the calculation.

The NLO renormalization constants of the gluon mass
and $g_s$ in $\overline{\rm MS}$ scheme read
\cite{Chetyrkin:1997fm,Bieri:2003ue}
\begin{eqnarray}
&& \delta Z_{x}^{(1),N_f} = -\frac{\alpha_s}{2\pi\epsilon}N_f,
\qquad\qquad
\delta Z_{g_s}^{(1),N_f} = \frac{\alpha_s}{6\pi\epsilon}N_f T_R
\qquad \mbox{with }\; T_R=\frac12 .
\end{eqnarray}
For the NNLO calculation we need NLO diagrams with counterterms, so that
the  full-theory  NLO diagrams are needed
up to order $\mathcal O(\epsilon)$.
For this reason we have extended the calculation
of Ref. \cite{NiersteNLO} to order $\epsilon^1$ for $m_c=0$.
Since the two-loop counterterms have $1/\epsilon^2$ poles, we further need
the full-theory  LO diagrams to order $\epsilon^2$. The results of these
diagrams can be found in Appendix~\ref{appa}.

The NNLO-large-$N_f$ piece of the
field renormalization constant for the external quark lines is
\begin{eqnarray}
&& \delta Z_{q}^{(2),N_f} =
\frac{\alpha_s^2}{(4\pi)^2}\frac{4}{3\epsilon}N_f,~~~ q=b,s
\end{eqnarray}
in the 't Hooft-Feynman gauge.

We now turn to the counterterms for the $\Delta
B=1$ operators.  The hamiltonian in \eq{Heff}
reads
\begin{eqnarray}
H^{\Delta
B=1}_{eff}&=&
\frac{G_F}{\sqrt{2}}V^*_{cs}V_{cb}
\sum^6_{j}\left[ C_j O_j \right]^{\rm bare} \,=\,
\frac{G_F}{\sqrt{2}}V^*_{cs}V_{cb}
\sum^6_{j}\left[ C_j O_j \right]^{\rm ren} \nn
&=&
\frac{G_F}{\sqrt{2}}V^*_{cs}V_{cb}
\sum^6_{j,k} C_j^{\rm bare} Z_{jk} O_k^{\rm ren}
\,=\,
\frac{G_F}{\sqrt{2}}V^*_{cs}V_{cb}
\sum^6_{j,k} C_j^{\rm ren} Z_{jk} O_k^{\rm bare}.  \label{eq:hren}
\end{eqnarray}
The last lines illustrates that one can view $Z_{jk}$ as
either renormalising the operator $O_k$ or the Wilson coefficient
$C_j$. Traditionally the renormalization is attributed to the
operator, but we adopt the latter viewpoint, with
$C_j\equiv C_j^{\rm ren}$ and $O_k\equiv O_k^{\rm bare}$.

Writing $Z_{jk}=\delta_{jk} + \delta Z_{jk}$
and  expanding $\delta Z_{jk}=\frac{\alpha_s}{4\pi} \delta
Z_{jk}^{(1)}+ \lt( \frac{\alpha_s}{4\pi}\rt)^2 \delta Z_{jk}^{(2)}+{\cal
  O}(\alpha_s^3)$ we find the following counterterms (first calculated
in Ref.~\cite{bw}) at order $\alpha_s^2 N_f$:
\begin{eqnarray}
&& \delta Z_{11}^{(2),N_f} = \delta Z_{22}^{(2),N_f} =
-\frac{1}{3}\delta Z_{12}^{(2),N_f} = -\left(\frac{1}{3\epsilon^2} +
\frac{1}{18\epsilon}\right)N_f,
\end{eqnarray}
which enters the result for $\dg$ in combination with the LO
(one-loop) matrix element $M^{(0)}$ of the full theory given in
(\ref{M0}).

For the penguin-diagram contributions we need the counterterms
$\delta Z_{2k}$ related to the mixing of $O_{2}$ into the
four-fermion operators $O_{3-6}$, necessary to renormalize the
penguin diagram $D_{11}$. There are two types of contributions. The
first type induces the mixing between $O_2$ and $O_{3-6}$. The
non-zero contributions are:
\begin{eqnarray}
&& \delta Z_{42}^{(1)} = \delta Z_{62}^{(1)} = \frac{1}{3\epsilon},
\\
&& \delta Z_{32}^{(2),N_f} = \delta Z_{52}^{(2),N_f} =
-\frac{2}{27\epsilon^2}N_f,
\\
&& \delta Z_{42}^{(2),N_f} = \delta Z_{62}^{(2),N_f} =
\frac{2}{9\epsilon^2}N_f.
\end{eqnarray}
In the result for $\dg$ the counterterms in the first line multiply
the matrix elements $M^{(1)}_{42}$ and $M^{(1)}_{62}$ in
\eq{eq:m2426}, while the other (two-loop) counterterms multiply
$M^{(0)}_{i2}$,
$i=3,\ldots,6$, in \eq{eq:m2i}. 
The second type of counterterms involves the
mixing of the penguin operators $O_{3-6}$ among themselves.
Together with $\delta Z_{42}^{(1)}$ and $\delta Z_{62}^{(1)}$
written above, the additional non-zero contributions, which multiply
the $M^{(0)}_{ij}$, $i,j=3,\ldots,6$,  in \eq{QCDpeng}, are:
\begin{eqnarray}
&& \delta Z_{32}^{(1)} = \delta Z_{52}^{(1)} = -\frac{1}{9\epsilon}.
\end{eqnarray}
Finally we state the $\mathcal O(\alpha_s)$ counterterms needed to
renormalize the penguin diagram $D_{12}$. Here the counterterms are
$\delta Z_{42}^{(1)}$ and $\delta Z_{62}^{(1)}$ noted above.

In the effective theory the counterterms for gluon mass, strong coupling
constant $g_s$, and external fields ($b$ and $s$) are treated as in the
full theory.  For the counterterms of the $\Delta B=2$ operators note
that here only the NNLO renormalization constants can contain parts
proportional to $N_f$, while the NLO renormalization constants have no pieces
proportional to $N_f$. Thus the $\overline{\rm MS}$ renormalization
of the $\Delta B=2$ operators at order $\alpha_s^2 N_f$
is trivial, one just has to drop the divergence from the considered
two-loop diagrams with quark loop.

\section{\bf Results for the coefficients $G$, $G_S$ at order
 $\alpha_s^2 N_f$\label{sec:cf}}
We first discuss the contributions $F$,$F_S$ to $G$,$G_S$  with two
insertions of $O_{1,2}$ (see \eq{deffp}).
We decompose $F$ defined as
\begin{eqnarray}
&& F (z) = F_{11}(z)C_1^2(\mu_1) + F_{12}(z)C_1(\mu_1)C_2(\mu_1) +
F_{22}(z)C_2^2(\mu_1),
\end{eqnarray}
with an analogous definition of $F_{S,ij}$.  We further write
\begin{eqnarray}
&& \nonumber F_{ij} (z) = F_{ij}^{(0)}(z) +
\frac{\alpha_s(\mu_1)}{4\pi}F_{ij}^{(1)}(z) +
\frac{\alpha_s^2(\mu_1)}{(4\pi)^2} \left(N_H F_{ij}^{(2),N_H}(1) +
N_V F_{ij}^{(2),N_V}(z) + N_L F_{ij}^{(2),N_L}(0)\right)
\end{eqnarray}
and similarly for $F_S(z)$.  $N_{H,V,L}$ are defined after \eq{eq:al2}.
The argument of $F_{ij}^{(2),N_{H,V,L}}$ is the ratio $z_q=m_q^2/m_b^2$,
where $m_q$ is the mass of the quark running in the loop in the gluon
propagator, i.e.\ $z_q$ equals 1,$z$, or 0.

The NNLO functions $F^{(2),N_f}_{ij}$ and $F^{(2),N_f}_{S,ij}$ for the
$b$ quark loop read:
\begin{eqnarray}
&& \nonumber F^{(2),N_H}_{11}(1) = -\frac{386}{9} \log
\frac{\mu_1}{m_b} +\frac{176}{9} \log \frac{\mu_2}{m_b} -\frac{40}{3}
\log \frac{\mu_1}{m_b} \log \frac{\mu_2}{m_b}+\frac{20}{3} \log ^2
\frac{\mu_2}{m_b}
\\
&& \hspace{2cm} +\pi ^2 \left(-\frac{2}{9} \left(1+104
\sqrt{5}\right)-\frac{64}{3}
\log\frac{1+\sqrt{5}}{2}\right)+\frac{64 \zeta
(3)}{3}+\frac{95993}{162},
\\
&& \nonumber F^{(2),N_H}_{12}(1) = \frac{554}{27}\log
\frac{\mu_1}{m_b}+\frac{352}{27} \log \frac{\mu_2}{m_b}
-\frac{80}{9} \log \frac{\mu_1}{m_b} \log
\frac{\mu_2}{m_b}+\frac{68}{3} \log ^2\frac{\mu_1}{m_b}+\frac{40}{9}
\log ^2\frac{\mu_2}{m_b}
\\
&& \hspace{2cm} +\pi ^2\left(-\frac{2}{27} \left(53+208
\sqrt{5}\right)-\frac{128}{9} \log
\frac{1+\sqrt{5}}{2}\right)+\frac{128 \zeta
(3)}{9}+\frac{518521}{1215},
\\
&& \nonumber F^{(2),N_H}_{22}(1) = \frac{236}{27} \log
\frac{\mu_1}{m_b}+\frac{58}{27}
   \log \frac{\mu_2}{m_b}-\frac{32}{9} \log \frac{\mu_1}{m_b} \log \frac{\mu_2}{m_b}+\frac{20}{3} \log^2\frac{\mu_1}{m_b}+\frac{16}{9} \log ^2\frac{\mu_2}{m_b}
\\
&& \hspace{2cm} +\pi ^2
   \left(\frac{4}{9} \log \frac{\mu_1}{m_b}-\frac{5}{27} \left(12+13\sqrt{5}\right)-\frac{20}{9} \log
\frac{1+\sqrt{5}}{2}\right)+\frac{14 \zeta
(3)}{9}+\frac{99511}{1215},
\\
&& \nonumber F^{(2),N_H}_{S,11}(1) = -\frac{80}{9}
\log\frac{\mu_1}{m_b}+\frac{320}{9} \log
\frac{\mu_2}{m_b}+\frac{128}{3} \log \frac{\mu_1}{m_b} \log
\frac{\mu_2}{m_b}-\frac{64}{3} \log ^2\frac{\mu_2}{m_b}
\\
&& \hspace{2cm} +\pi ^2 \left(-\frac{16}{9} \left(8+13
   \sqrt{5}\right)-\frac{64}{3} \log \frac{1+\sqrt{5}}{2}\right)+\frac{64\zeta (3)}{3} + \frac{295238}{405},
\\
&& \nonumber F^{(2),N_H}_{S,12}(1) = \frac{464}{27} \log
\frac{\mu_1}{m_b}+\frac{640}{27} \log \frac{\mu_2}{m_b}+\frac{256}{9}
\log \frac{\mu_1}{m_b} \log \frac{\mu_2}{m_b}+\frac{32}{3} \log
   ^2\frac{\mu_1}{m_b}-\frac{128}{9} \log^2\frac{\mu_2}{m_b}
\\
&& \hspace{2cm} +\pi ^2 \left(-\frac{16}{27} \left(19+26
\sqrt{5}\right)-\frac{128}{9} \log
\frac{1+\sqrt{5}}{2}\right)+\frac{128\zeta(3)}{9} + \frac{121724}{243},
\\
&& \nonumber F^{(2),N_H}_{S,22}(1) = \frac{704}{27} \log
\frac{\mu_1}{m_b}-\frac{320}{27}
   \log \frac{\mu_2}{m_b}-\frac{128}{9} \log \frac{\mu_1}{m_b} \log \frac{\mu_2}{m_b}+\frac{32}{3} \log
   ^2\frac{\mu_1}{m_b}+\frac{64}{9} \log ^2\frac{\mu_2}{m_b}
\\
&& \hspace{2cm} +\pi ^2
   \left(-\frac{32}{9} \log \frac{\mu_1}{m_b}+\frac{8}{27} \left(30-13 \sqrt{5}\right)-\frac{32}{9} \log
   \frac{1+\sqrt{5}}{2} \right)+\frac{80 \zeta (3)}{9}+\frac{5836}{1215}.
\end{eqnarray}
The result for the charm loop quark is expanded in $z=m_c^2/m_b^2$
up to $\mathcal O(z^3)$:
\begin{eqnarray}
\label{resz1} && \nonumber F^{(2),N_V}_{11}(z) = -42.8889 \log
   \frac{\mu_1}{m_b}+19.5556 \log\frac{\mu_2}{m_b}-13.3333 \log \frac{\mu_1}{m_b} \log \frac{\mu_2}{m_b}+6.66667 \log ^2\frac{\mu_2}{m_b}
\\
&&  \nonumber \hspace{2cm} -5.84736-39.4784\sqrt{z}+z (37-24\log z)-39.4784z^{3/2}
\\
&&  \nonumber \hspace{2cm} + z^2 (2\log^2 z-63.5556 \log z+24.5336)
\\
&& \hspace{2cm} +z^3 \left( -14.2222\log ^2 z+35.8963 \log z+69.8579\right) + \mathcal O(z^4),
\\
\label{resz2} && \nonumber F^{(2),N_V}_{12}(z) = 20.5185 \log
\frac{\mu_1}{m_b}+13.037 \log \frac{\mu_2}{m_b}-8.88889 \log
\frac{\mu_1}{m_b} \log \frac{\mu_2}{m_b}+22.6667 \log^2\frac{\mu_1}{m_b}
\\
&&  \nonumber \hspace{2cm} +4.44444 \log^2\frac{\mu_2}{m_b} +
40.0184-26.3189\sqrt{z}-z (16 \log z+111.333)-26.3189z^{3/2}
\\
&& \nonumber \hspace{2cm} +z^2 \left(18.3333 \log ^2 z-117.926 \log
z+86.7372\right)
\\
&& \hspace{2cm}  + z^3 \left(-9.48148 \log ^2 z+20.9086 \log z +
62.3882\right) + \mathcal O(z^4),
\\
\label{resz3} && \nonumber F^{(2),N_V}_{22}(z) =  13.1272 \log
\frac{\mu_1}{m_b} + 2.14815 \log \frac{\mu_2}{m_b}-3.55556 \log
\frac{\mu_1}{m_b} \log \frac{\mu_2}{m_b} + 6.66667
\log^2\frac{\mu_1}{m_b}
\\
&& \nonumber \hspace{2cm}   + 1.77778 \log
   ^2 \frac{\mu_2}{m_b} + 20.858 - 52.6379 \sqrt{z} - z (18.1739+32 \log z) + 35.0919 z^{3/2}
\\
&& \nonumber \hspace{2cm} + z^2 \left(-2.83333 \log ^2 z-16.6481 \log
z+13.9138\right)
\\
&& \hspace{2cm} + z^3 \left(-1.48148 \log ^2 z+9.29383 \log
z+0.204084\right) + \mathcal O(z^4),
\\
\label{resz4} && \nonumber F^{(2),N_V}_{S,11}(z) = -8.88889 \log
\frac{\mu_1}{m_b} + 35.5556 \log \frac{\mu_2}{m_b} + 42.6667 \log
\frac{\mu_1}{m_b} \log \frac{\mu_2}{m_b} - 21.3333 \log ^2
\frac{\mu_2}{m_b}
\\
&& \nonumber \hspace{2cm} +82.4693-157.914\sqrt{z} +136 z-157.914z^{3/2}
\\
&& \nonumber \hspace{2cm} + z^2 (8\log^2 z-75.5556 \log z+75.1571)
\\
&& \hspace{2cm} + z^3 \left(-14.2222 \log^2 z+39.2296 \log
z+68.3912\right) + \mathcal O(z^4),
\\
\label{resz5} && \nonumber F^{(2),N_V}_{S,12}(z) = 17.1852 \log
\frac{\mu_1}{m_b} + 23.7037 \log \frac{\mu_2}{m_b} + 28.4444 \log
\frac{\mu_1}{m_b} \log \frac{\mu_2}{m_b} + 10.6667
\log^2\frac{\mu_1}{m_b}
\\
&& \nonumber \hspace{2cm} - 14.2222 \log^2 \frac{\mu_2}{m_b} + 75.6462 -105.276\sqrt{z} + 26.6667 z -105.276 z^{3/2}
\\
&& \nonumber \hspace{2cm} + z^2 \left(13.3333 \log ^2 z-85.9259 \log
z+83.2254\right)
\\
&& \hspace{2cm} + z^3 \left(-9.48148 \log ^2 z + 24.7309 \log z +
53.0371\right) + \mathcal O(z^4),
\\
\label{resz6} && \nonumber F^{(2),N_V}_{S,22}(z) = - 9.01785 \log \frac{\mu_1}{m_b} - 11.8519 \log \frac{\mu_2}{m_b} -14.2222 \log
\frac{\mu_1}{m_b} \log \frac{\mu_2}{m_b} + 10.6667 \log^2\frac{\mu_1}{m_b}
\\
&& \nonumber \hspace{2cm}  + 7.11111 \log^2\frac{\mu_2}{m_b} -
42.0084 + 105.276 \sqrt{z} - 174.609 z + 666.747 z^{3/2}
\\
&& \nonumber \hspace{2cm}  + z^2 \left(-57.3333 \log
   ^2 z+236.296 \log z - 526.684\right)
\\
&& \hspace{2cm} + z^3 \left(-2.37037 \log ^2 z + 28.5235 \log z -
32.2992\right) + \mathcal O(z^4).
\end{eqnarray}
The contribution  of each light quark $u,d,s$ can be obtained by
setting $z=0$ in \eqsto{resz1}{resz6}, i.e.\
$F_{ij}^{(2),N_L}(0)=F_{ij}^{(2),N_V}(0)$.

For the contributions of penguin diagrams and penguin operators
in \eq{deffp} we write
\begin{eqnarray}
P(z) &=& P^{\rm NLO} (z) + \Delta P^{\rm NNLO}(z),\qquad
P_S(z) \;=\; P_S^{\rm NLO}(z) + \Delta P_S^{\rm NNLO} (z),
\end{eqnarray}
where $P^{\rm NLO}(z)$ and $P_S^{\rm NLO}(z)$ are the NLO results of
Ref.~\cite{NiersteNLO}, while $\Delta P(z)$ and $\Delta P_S(z)$ are the
NNLO corrections with $z$ . Since we treat $C_{3-6}$ as ${\cal
  O}(\alpha_s)$, the latter contain terms of order $C_{3-6}C_{3-6}$,
$\alpha_s C_2C_{3-6}$, and terms of order $\alpha_s^2C_2^2$. The
large-$N_f$ part of $\Delta P^{\rm NNLO}(z)$  is decomposed as
\begin{eqnarray}
\Delta P^{\rm NNLO}(z)=N_H \Delta P^{{\rm NNLO},N_H}(1)+
N_V \Delta P^{{\rm NNLO},N_V}(z)+N_L \Delta P^{{\rm NNLO},N_L}(0), \no
\end{eqnarray}
with an analogous formula for $\Delta P_S^{\rm NNLO}(z)$.  In the
penguin contributions the charm mass on all lines touching $O_2$ are set
to zero, while all other charm loops are kept massive.  These include
not only the loop in $D_{11-13}$, but also the loops connecting two
penguin operators $O_{3-6}$ or one penguin operator and a charm-gluon
vertex. The latter two contributions appear in counterterm diagrams (to
e.g.\ $D_{11-13}$) and must be treated in the same way as the diagrams
which they renormalize.  Consequently, the argument $z_q$ (with
$z_q=1,z,$ or 0) in $\Delta P^{{\rm NNLO},N_{H,V,L}}(z_q)$ refers to the mass
in the loop of any of these three situations. (At NNLO there are no
diagrams with more than one loop.)

The results are:
\begin{eqnarray}
\Delta P^{{\rm NNLO},N_H}(1) &=& \frac{\alpha_s(\mu_1)}{4\pi}
G_p^{(1),N_H}(1) M_4'(\mu_1) + \frac{\alpha_s^2(\mu_1)}{(4\pi)^2}
G_p^{(2),N_H}(1) C_2^2(\mu_1),
\\
\Delta P_S^{{\rm NNLO},N_H}(1) &=& -\frac{\alpha_s(\mu_1)}{4\pi} 8
G_p^{(1),N_H}(1) M_4'(\mu_1) - \frac{\alpha_s^2(\mu_1)}{(4\pi)^2} 8
G_p^{(2),N_H}(1) C_2^2(\mu_1),
\end{eqnarray}
\begin{eqnarray}
\Delta P^{{\rm NNLO},N_V}(z) &=& \sqrt{1-4z}\left((1-z)M_1'(\mu_1)+
\frac{1}{2}(1-4z)M_2'(\mu_1)+3zM_3'(\mu_1)\right) \nonumber
\\
&&  
+ \frac{\alpha_s(\mu_1)}{4\pi}
G_p^{(1),N_V}(z) M_4'(\mu_1) + \frac{\alpha_s^2(\mu_1)}{(4\pi)^2}
G_p^{(2),N_V}(z) C_2^2(\mu_1), \label{pengz1}
\\
\Delta P_S^{{\rm NNLO},N_V}(z) &=&
\sqrt{1-4z}\,(1+2z)\left(M_1'(\mu_1)-M_2'(\mu_1)\right) \nonumber
\\
&&  
- \frac{\alpha_s(\mu_1)}{4\pi}
8G_p^{(1),N_V}(z) M_4'(\mu_1) - \frac{\alpha_s^2(\mu_1)}{(4\pi)^2}
8G_p^{(2),N_V}(z) C_2^2(\mu_1),\label{pengz2}
\end{eqnarray}
with
\begin{eqnarray}
&& G_p^{(1),N_H}(1) = -\frac{1}{54}\left(6\log\frac{\mu_1}{m_b} -
3\sqrt{3}\pi+17\right),
\\
&& G_p^{(2),N_H}(1) = \frac{2}{81}\left(6\log\frac{\mu_1}{m_b} -
3\sqrt{3}\pi+17\right) \left[2\log\frac{\mu_1}{m_b} + \frac{2}{3} +
\frac{3C_8(\mu_1)}{C_2(\mu_1)}\right],
\end{eqnarray}
\begin{eqnarray}
&&  G_p^{(1),N_V}(z) =  - \frac{1}{54}
\left[\sqrt{1-4z}(1+2z)\left(6\log\frac{\mu_1}{m_b}+3\log\sigma+2\right)+6\log\frac{\mu_1}{m_b}
-3\log z +5+12z \right. \nonumber
\\
&& \left. \hspace{2.4cm}+ \frac{9C_8(\mu_1)}{C_2(\mu_1)}
\sqrt{1-4z}\,(1+2z) \right],
\\
\label{pengz4} && \nonumber G_p^{(2),N_V}(z) =
\frac{1}{81}\left[\frac{4}{3}\left(3\log\frac{\mu_1}{m_b}+1\right)\left(\sqrt{1-4z}\,(1+2z)\left(3\log\frac{\mu_1}{m_b}+3\log
\sigma+1\right) + 6\log\frac{{\mu_1}}{m_b} \right.\right.
\\
&& \nonumber \hspace{2.5cm} \left. -3\log z +5+12z
\right)-3\pi^2\sqrt{1-4z}\, (1+2z)
\\
&& \nonumber \hspace{1.2cm} + \frac{6 C_8(\mu_1)}{C_2(\mu_1)}
\left(\sqrt{1-4z}(1+2z)\left(6\log\frac{\mu_1}{m_b}+3\log\sigma+2\right)
+6\log\frac{\mu_1}{m_b} -3\log z + 5 + 12z\right.
\\
&& \left.\left. \hspace{2.2cm} + \frac{9C_8(\mu_1)}{2C_2(\mu_1)}
\sqrt{1-4z}(1+2z)\right)\right],
\end{eqnarray}
where we have defined $M_1'=3C_3^2+2C_3C_4+3C_5^2+2C_5C_6$,
$M_2'=C_4^2+C_6^2$, $M_3'=2(3C_3C_5+C_3C_6+C_4C_5+C_4C_6)$,
$M_4'=2(C_2C_4+C_2C_6)$ and
\begin{eqnarray}
\sigma=\frac{1-\sqrt{1-4z}}{1+\sqrt{1-4z}}.
\end{eqnarray}
As above, $\Delta P^{{\rm NNLO},N_L}(0)$
is obtained from  \eqsand{pengz1}{pengz2} by setting $z$ to 0,
i.e.\ $\Delta P^{{\rm NNLO},N_L}(0)=\Delta P^{{\rm NNLO},N_V}(0)$.


In the matching procedure one has to take into account that the
operators, couplings and masses on the full-theory side are defined at
the scale $\mu_1$, while the effective operators are defined at the
scale $\mu_2$. To compare both sides one must choose the same expansion
parameter on both sides, e.g.\ $\alpha_s(\mu_1)$, and use \eq{scale} for
this. Therefore the $\alpha_s^2 N_f$ results quoted in this section also
contain contributions from the $\alpha_s^1$ parts through \eq{scale}.

\section{\bf Phenomenology   of $\Delta\Gamma$\label{sec:num}}
In this section we show the impact of the new $\alpha_s^2 N_f$ terms
on $\Delta\Gamma_s$. Our input parameters are collected in
\tab{tab:inp}. We use the complete NNLO $\Delta B=1$ Wilson
coefficients $C_1$, $C_2$ \cite{Buras:2006gb} and the complete NLO
expressions for $C_3,...C_6$, with the numerical values listed in
\tab{tab:Wilson}. The $\alpha_s^2 N_f^0$ terms of the coefficients
inflict a scheme dependence on $\dg$, which will only be cancelled
once the full NNLO calculation is performed. Nevertheless we can
study whether the new large-$N_f$ terms help to reduce scale and
scheme dependences.

The coefficients $G=F+P$ and $G_S=-F_S-P_S$ correspond to the pole
scheme for $\dg$. For the $\ov{\rm MS}$ scheme we must multiply these
coefficients with $\bar m_b^2/m_b^{\rm pole}$ and expand this ratio to
the order in $\alpha_s$ to which $G,G_s$ are calculated
\cite{NiersteNLONB}, in our case this is ${\cal O}(\alpha_s^2 N_f)$.  In
both schemes we use $\bar z$ defined in \eq{eq:defz}; the transformation
from $z$ to $\bar z$ in the NLO formula can be found in Eq.~(18) of
Ref.~\cite{Beneke:2002rj}. Since we have set $z=0$ in the charm lines
attached to weak vertices, no NNLO corrections to the transformation
occur.

We further must calculate $m_b^{\rm pole}$ from $\bar m_b$ and we use
the full 2-loop result for this
\cite{Gray:1990yh,Chetyrkin:1999qi,Asatrian:2006rq}. This is a
reasonable approach, if the missing $\alpha_s^2 N_f^0$ in the $\ov{\rm
  MS}$ scheme have the expected ${\cal O}(10\%)$ size while being
larger in the pole scheme to compensate for the anomalously large ratio
$m_b^{\rm pole\,2}/\bar m_b^2\sim 1.3$.

In both $\ov{\rm MS}$ and pole scheme we use $\bar m_b(\bar m_b)= (4.18
\pm 0.03)\,\gev$ as input and calculate $m_b^{\rm pole}=4.58\gev$ at NLO
and $m_b^{\rm pole}=4.85\gev$ at order $\alpha_s^2$.
\begin{table}[t]
{\begin{displaymath}
\begin{array}{rll@{~~}rll}
\bar m_b(\bar m_b)=& (4.18 \pm 0.03)\,\gev
&\mbox{\cite{Agashe:2014kda}} & \bar m_c(\bar m_c)=& (1.286 \pm
0.013_{\rm stat}
                      \pm 0.040_{\rm syst})\,\gev &
          \mbox{\cite{Charles:2004jd,Kuhn:2007vp,Allison:2008xk}} \\
\bar m_s(\bar m_b)=& (0.079\pm 0.002)\, \gev &
  \mbox{\cite{Aoki:2016frl,Bazavov:2016nty}} &
\bar m_t (m_t)= & (165.96\pm 0.35_{\rm stat}  \pm 0.64_{\rm syst})\,
   \gev &  \mbox{\cite{Charles:2004jd}} \\
|V_{cb}| = &  41.80 \epm{0.33}{0.68}\cdot 10^{-3} &
\mbox{\cite{Charles:2004jd}} &
|V_{ub}| = &  3.714 \epm{0.07}{0.06}\cdot 10^{-3} &  \mbox{\cite{Charles:2004jd}}\\
\gamma =& 68^\circ \epm{0.9^\circ}{2.0^\circ} &
\mbox{\cite{Charles:2004jd}} &
m_b^{\rm pow} =& 4.7\, \gev & \mbox{see \cite{NiersteNLONB}}\\
f_{B_s} \sqrt{\widetilde B_S^\prime} =& 303 MeV & \mbox{\cite{Bazavov:2016nty}} &
\widetilde{B}_{R_0} =& 0.56 \pm 0.53 & \mbox{\cite{Bazavov:2016nty}} \\
f_{B_s} \sqrt{B}=& 224 MeV  & \mbox{\cite{Bazavov:2016nty}} &
\\
M_{B_s}=& 5.368\, \gev &\mbox{\cite{Agashe:2014kda}} & &
\alpha_s(M_Z)= 0.1185 & \\
|V_{ts}^* V_{tb}|=& 40.9\cdot 10^{-3} &&& \\
\end{array}
\end{displaymath}}
\caption{Input parameters used in Sec.~\ref{sec:num}.
$\bar m_s(\bar m_b)$ is calculated from $\bar m_s(2\gev)=0.094\pm
0.001\,\gev$ \cite{Aoki:2016frl}.
 $m_B^{\rm pow}$ is a redundant parameter
  calibrating the overall size of the hadronic parameters $B_{R_i}$ which
quantify the matrix elements at order $\lqcd/m_b$. The translation
of $\langle R_0\rangle= -0.19\pm 0.18\,\gev$ \cite{Bazavov:2016nty}
to $\widetilde{B}_{R_0}$ for our choice of $m_b^{\rm pow}$ is done
with $f_{B_s}=(0.224\pm 0.05)\,\gev$ \cite{Aoki:2016frl}.
Subsequently this result is used to rescale
$\widetilde{B}_{R_0}/B $ from the value in
Ref.~\cite{Bazavov:2016nty} to the one in the table.
$\widetilde{B}_S$ is larger than $\widetilde{B}_S^\prime$ by a
factor of $M_{B_s}^2/(\bar m_b+\bar m_s)^2=1.588$, so that
$\widetilde{B}_S^\prime/B=1.83\pm 0.21$. \label{tab:inp}}
~\\[-5mm]\hrule
\end{table}
\begin{table}[tb]
\begin{center}
\begin{tabular}{|r|r|r|r|}
\hline $i$ & $C_i^{(0)}(\mu_b)$ & $C_i^{(1)}(\mu_b)$  &
$C_i^{(2)}(\mu_b)$ \\\hline
1 & $-$0.2687 &   4.332 & 50.142\\
2 &    1.1179 & $-$2.024 & $-$17.114\\
3 &    0.0121 &    0.090 & $-$~~~~ \\
4 & $-$0.0274 & $-$0.465 & $-$~~~~ \\
5 &    0.0079 &    0.041 & $-$~~~~ \\
6 & $-$0.0343 & $-$0.434 & $-$~~~~ \\
8 & $-$0.1508 & $-$1.0006 & $-$~~~~ \\
\hline
\end{tabular}
\caption{The LO, NLO and NNLO Wilson
coefficients~$C_i^{(k)}(\mu_b)$~ at $\mu_b=\bar{m}_b = 4.18\;{\rm
GeV}$ using $\alpha_s(\bar{m}_b)=0.226$ (implementing the formula of
Ref.~\cite{Huber:2005ig} with QED effects set to zero) and
the matching scale $\mu_0=M_W$.
We have used Ref.~ \cite{Buras:2006gb} to compute $C_1^{(k)}(\mu_b)$ and
$C_2^{(k)}(\mu_b)$. The NLO piece of the Wilson coefficient $C_8^{(1)}$
is taken from the calculation in a different basis \cite{Gorbahn:2005sa}
and the quoted value therefore neglects a numerically small contribution
from an evanescent operator. \label{tab:Wilson}}
\end{center}
~\\[-5mm]\hrule
\end{table}
\begin{figure}[t]
\begin{center}
\vspace{1.0cm} \hspace{0.5cm}
\includegraphics[scale=0.75]{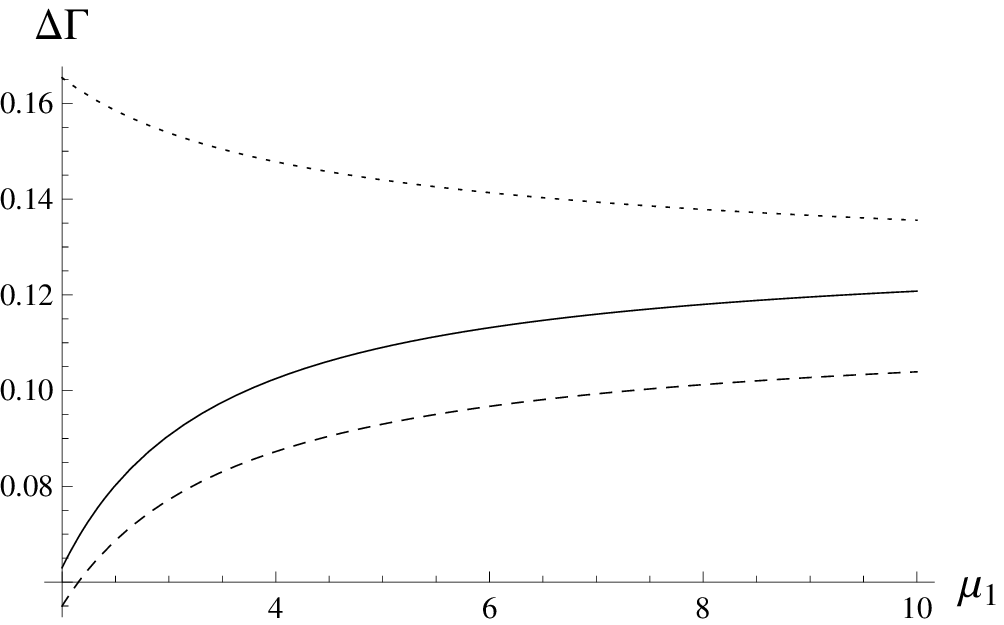}
\hspace{0.5cm}
\includegraphics[scale=0.75]{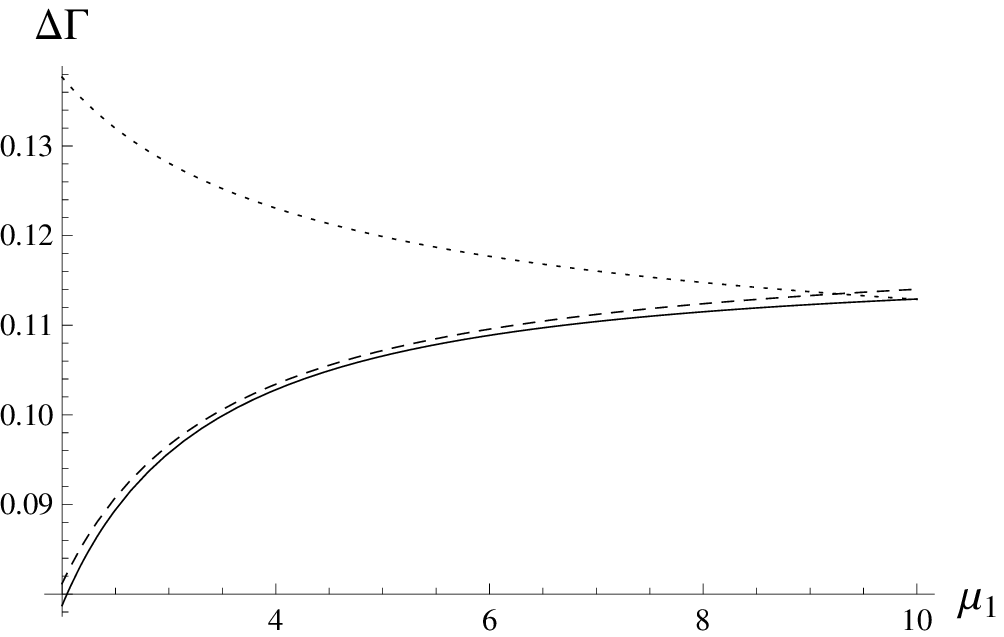}
 \end{center}
 \caption{Renormalization scale dependence for $\Delta\Gamma$ at LO
   (dotted), NLO (dashed), and NNLO (solid) results for the pole scheme
   (left) and the $\ov{\rm MS}$ scheme (right). On the x axis is $\mu_1$ in GeV-s. 
 }
 \label{fig2}
~\\[-2mm]\hrule
\end{figure}
In \eq{eq:numintro} we find a small scale dependence in the $\ov{\rm
  MS}$ scheme, because the sizable $\mu_1$ dependence of
the prefactor $\bar m_b(\mu_1)^2$ cancels nicely with the $\mu_1$
dependence of $G$,$G_S$. In our partial NNLO result this efficient
cancellation is less pronounced than in the NLO result of \eq{eq:numintro}.
To be conservative, we therefore use a different approach
in this section: We keep $\bar m_b(\bar m_b)^2$ fixed and, for
consistency, also eliminate the $\log (\mu_1/\bar m_b)$ terms related to
the running of $\bar m_b$ from $G$,$G_S$. This leads to a larger $\mu_1$
dependence at NLO.

We find:
\begin{align}
\dg^{NLO} &= \lt( 0.091 \pm 0.020_{\rm scale} \rt) \, \gev \qquad
\mbox{(pole)} \nn \dg^{NLO} &= \lt( 0.104 \pm 0.015_{\rm
scale}\rt)\, \gev \qquad \mbox{($\ov{\rm MS}$)} \label{eq:numres1}
\\[2mm]
\dg^{NNLO} &= \lt( 0.108 \pm 0.021_{\rm scale} \rt) \, \gev \qquad
\mbox{(pole)} \nn \dg^{NNLO} &= \lt( 0.103 \pm 0.015_{\rm scale}\rt)\,
\gev \qquad \mbox{($\ov{\rm MS}$)} \label{eq:numres2}
\end{align}
where the scale dependence is calculated by varying $\mu_1$ between
 $m_b/2$ and $2m_b$ and
for the quoted central values of $\Delta\Gamma$ we took
$\mu_1=m_b^{\rm pole}$ and
$\mu_1=\bar{m_b}$ for the pole and $\ov{\rm MS}$ schemes, respectively.
Unlike in \eq{eq:numintro} other sources of error are neglected here.
The $\mu_1$ dependence is plotted in \fig{fig2}.

We observe that the partial NNLO corrections calculated in this section
decrease the scheme dependence and give preference to the NLO result in
the $\ov{\rm MS}$ scheme. The result also suggests that in
\eq{eq:numintro} the $\mu_1$ dependence is underestimated and that the
partial NNLO calculation does not reduce the scale dependence to a
satisfactory level.

We have discussed the naive non-abelianization approach (NNA)
in Sec.~\ref{sec:tf}. If we trade $N_f$ for $\beta_0$
in $G$, $G_S$ and the relation between $\bar{m}_b=4.18\,\gev$
and $m_b^{\rm pole}$, we find $m_b^{\rm pole}=4.87\,\gev$, which is close
to the full two-loop result, and
\begin{align}
\dg^{NNA} &= \lt( 0.071 \pm 0.020_{\rm scale} \rt) \, \gev \qquad
\mbox{(pole)} \nn
\dg^{NNA} &= \lt( 0.099 \pm 0.012_{\rm scale}\rt)\,
\gev \qquad \mbox{($\ov{\rm MS}$)}. \label{eq:nnanum}
\end{align}
Comparing \eq{eq:numres2} with \eq{eq:nnanum} we find that the
$\ov{\rm MS}$ result is quite stable, if we change the literal
$\alpha_s^2N_f$ result to the NNA one, while the pole-scheme result is
not.

Until a full NNLO calculation is available, we recommend to use the
$\ov{\rm MS}$ NLO value with an enlarged $\mu_1$ dependence compared to
\eqsand{eq:dgb}{eq:numintro}:
\begin{align} 
\dg &=\;  (1.86\pm 0.17) \,
          f_{B_s}^2 B  \; +\;  (0.42\pm 0.03) f_{B_s}^2 \widetilde B_S^\prime
         \; +\; (-0.55 \pm 0.29) \, f_{B_s}^2. \nn
\dg &= \lt( 0.104 \pm 0.015_{\rm scale} \pm
   0.007_{B,\widetilde B_S}
   \pm 0.015_{\lqcd/m_b} \rt)\, \gev
\qquad \mbox{($\ov{\rm MS}$)} \label{eq:fin}
\end{align}

\section{Conclusions}
We have calculated the contributions of order $\alpha_s^2 N_f$ to the width
difference in the \bbs\ system in an expansion in $m_c/m_b$, neglecting
terms of order $(m_c/m_b)^2$ and higher.  This calculation has involved
three-loop massive master integrals with two mass scales.  We find a
larger correction for the decay width difference in the pole scheme and
only a minuscule correction for the $\ov{\rm MS}$ scheme.  As a result,
the scheme dependence reduces considerably and we advocate the use of
the NLO numerical values in \eq{eq:fin}.

\section*{Acknowledgments}
We would like to thank Christoph Greub for helpful discussions. This
work has been supported by Grant No. 86426 of the Volkswagen
Stiftung. H.A., A.H. and A.Y. were further supported by the State
Committee of Science of Armenia Program Grant No. 15T-1C161 and U.N.
has received support from BMBF under contract no.~05H15VKKB1.

\appendix
\section{Full-theory matrix elements\label{appa}}
In this section we collect the needed unrenormalized LO and NLO matrix elements
to order $\epsilon^2$ and $\epsilon$, respectively. We decompose the
matrix element as
\begin{equation}
M \,= \, M_{\rm cc} + M_{\rm peng}, \label{mccp}
\end{equation}
where the first term denotes the contribution with two insertions
of the current-current operators $O_{1,2}$ and the second term
comprises the diagrams with at least one penguin operator. Recall that
we count $C_{3-6}$ as order $\alpha_s$, so that one loop less is needed
for $M_{\rm peng}$ compared to $M_{\rm cc}$. We expand
$M_{\rm cc,peng}=M_{\rm cc,peng}^{(0)}+
        \frac{\alpha_s}{4\pi}M_{\rm cc,peng}^{(1)}+\ldots$
and quote all
results for $m_c=0$.

\subsection{Current-current operators}
The LO full-theory result $M^{(0)}_{\rm cc}$ is needed to order $\mathcal
O(\epsilon^2)$:
\begin{eqnarray}
\label{M0} && \nonumber M^{(0)}_{\rm cc} = -\frac{G^2_F m_b^2}{12\pi}
(V^*_{cs}V_{cb})^2\left(\left(3 C_1^{b2}+2C_1^b C_2^b\right)
\left(\frac{1}{2} \langle Q\rangle^{(0)} - \langle \tilde{Q}_S\rangle^{(0)}\right)
   +  C_2^{b2} \left(\langle Q\rangle^{(0)} +
\langle \tilde{Q}_S\rangle^{(0)}\right)\right)
\\
&& \hspace{1.5cm}
  \cdot \left(1+\epsilon  \left(\frac{2}{3}+2\log \frac{\mu_1}{m_b}\right)+\epsilon ^2 \left(2 \log ^2 \frac{\mu_1}{m_b}+\frac{4}{3} \log \frac{\mu_1}{m_b}-\frac{\pi^2}{4}+\frac{13}{9}\right)\right),
\end{eqnarray}
Here and in the following
$\langle ...\rangle^{(0)}$ denote tree-level matrix elements and
$C_k^b=\sum_j C_j Z_{jk}$ are bare Wilson coefficients
(see \eq{eq:hren}).

We decompose the NLO diagrams according to the diagrams in \fig{fig1}
and the Wilson coefficients as
\begin{eqnarray}
M^{(1)}_{\rm cc}&=&
-\frac{G^2_F m_b^2}{12\pi}
(V^*_{cs}V_{cb})^2 \lt(
M^{(1)}_{11,D_{1-10}} +
M^{(1)}_{12,D_{1-10}} +
M^{(1)}_{22,D_{1-10}} +
M^{(1)}_{D_{11}} +
M^{(1)}_{D_{12}} \rt)
\end{eqnarray}

The sum of the full-theory non-penguin NLO diagrams amounts to
\begin{eqnarray}
&& \nonumber M^{(1)}_{11,D_{1-10}} = C_1^{b2}
\left(\langle Q\rangle^{(0)}\left[-\frac{17}{3}-4\log \frac{\mu_1}{m_b}+4\log
\frac{\mu_1}{m_g}+\epsilon\left(\log\frac{\mu_1}{m_b}\left(8\log\frac{\mu_1}{m_g}-\frac{125}{6}\right)\right.\right.\right.
\\
&& \nonumber \left.\left.\left. \hspace{4cm}
-12\log^2\frac{\mu_1}{m_b}+4\log^2\frac{\mu_1}{m_g}-\frac{11}{6}\log\frac{\mu_1}{m_g}-48\zeta(3)+\frac{4\pi^2}{3}+\frac{565}{72}\right)\right]\right.
\\
&& \nonumber \left. \hspace{3.3cm}
+\langle \tilde{Q}_S\rangle^{(0)}\left[\frac{4}{3}-16\log\frac{\mu_1}{m_b}+16\log\frac{\mu_1}{m_g}+\epsilon\left(\log
\frac{\mu_1}{m_b}\left(32\log\frac{\mu_1}{m_g}-\frac{28}{3}\right)\right.\right.\right.
\\
&& \left.\left.\left. \hspace{3.7cm}
-48\log^2\frac{\mu_1}{m_b}+16\log^2\frac{\mu_1}{m_g}+\frac{44}{3}\log\frac{\mu_1}{m_g}+96\zeta
(3)+\frac{16\pi^2}{3}-\frac{671}{9}\right)\right]\right),
\\
&& \nonumber M^{(1)}_{12,D_{1-10}}
\\
&& \nonumber \hspace{1cm} = 2C_1^bC_2^b \left(\langle Q\rangle^{(0)}
\left[-\frac{13}{2\epsilon }-\frac{385}{18}-\frac{82}{3}
\log\frac{\mu_1}{m_b}+\frac{4}{3}\log\frac{\mu_1}{m_g}+\epsilon\left(\log\frac{\mu_1}{m_b}\left(\frac{8}{3}\log\frac{\mu_1}{m_g}-\frac{1529}{18}\right)\right.\right.\right.
\\
&& \nonumber \left.\left.\left. \hspace{4cm}
-56\log^2\frac{\mu_1}{m_b}+\frac{4}{3}\log^2\frac{\mu_1}{m_g}-\frac{11}{18}\log\frac{\mu_1}{m_g}-16\zeta(3)+\frac{133\pi^2}{36}-\frac{9263}{216}\right)\right]\right.
\\
&& \nonumber \left. \hspace{2.5cm} + \langle \tilde{Q}_S\rangle^{(0)}
\left[\frac{4}{\epsilon }+\frac{112}{9}+\frac{32}{3}\log
\frac{\mu_1}{m_b}+\frac{16}{3}\log\frac{\mu_1}{m_g}+\epsilon\left(\log\frac{\mu_1}{m_b}\left(\frac{32}{3}\log\frac{\mu_1}{m_g}+\frac{404}{9}\right)\right.\right.\right.
\\
&& \left.\left.\left. \hspace{4cm}
+16\log^2\frac{\mu_1}{m_b}+\frac{16}{3}\log^2\frac{\mu_1}{m_g}+\frac{44}{9}\log\frac{\mu_1}{m_g}+32\zeta(3)-\frac{2\pi^2}{9}+\frac{85}{27}\right)\right]\right),
\\
&& \nonumber M^{(1)}_{22,D_{1-10}}
\\
&& \nonumber \hspace{0.8cm} = C_2^{b2}
\left(\langle Q\rangle^{(0)}\left[-\frac{1}{\epsilon}-\frac{11}{18}-\frac{5}{3}\pi^2-\frac{8}{3}\log\frac{\mu_1}{m_b}-\frac{4}{3}\log\frac{\mu_1}{m_g}
+\epsilon\left(\log\frac{\mu_1}{m_b}\left(\frac{11}{18}-\frac{20\pi^2}{3}-\frac{8}{3}\log\frac{\mu_1}{m_g}\right)
\right.\right.\right.
\\
&& \nonumber \left.\left.\left. \hspace{4cm} -
4\log^2\frac{\mu_1}{m_b}-\frac{4}{3}\log
^2\frac{\mu_1}{m_g}-\frac{55}{18}\log\frac{\mu_1}{m_g}-22\zeta(3)-\frac{49\pi
^2}{18}+\frac{445}{24}\right)\right]\right.
\\
&& \nonumber \left. \hspace{0.6cm}
+\langle \tilde{Q}_S\rangle^{(0)}\left[\frac{8}{\epsilon}-\frac{8}{3}\pi^2+\frac{320}{9}+
\frac{112}{3}\log\frac{\mu_1}{m_b}-\frac{16}{3}
\log\frac{\mu_1}{m_g}+\epsilon\left(\log\frac{\mu_1}{m_b}\left(\frac{1324}{9}-\frac{32}{3}\pi^2-\frac{32}{3}\log\frac{\mu_1}{m_g}\right)\right.\right.\right.
\\
&& \left.\left.\left. \hspace{3.5cm} +80\log
^2\frac{\mu_1}{m_b}-\frac{16}{3}\log^2\frac{\mu_1}{m_g}-\frac{44}{9}\log\frac{\mu_1}{m_g}-16\zeta(3)-\frac{92\pi^2}{9}+\frac{3443}{27}\right)\right]\right).
\end{eqnarray}
and for the penguin diagrams
\begin{eqnarray}
&& \nonumber M^{(1)}_{D_{11}} =
-\frac{5\langle Q\rangle^{(0)}+8\langle \tilde{Q}_S\rangle^{(0)}}{9}
  C_{b2}^2\left(\frac{1}{\epsilon}+\frac{4}{3}
+4\log
\frac{\mu_1}{m_b}+\epsilon\left(\frac{10}{3}-\frac{5}{6}\pi^2+\frac{16}{3}\log
\frac{\mu_1}{m_b}+8\log^2 \frac{\mu_1}{m_b}\right)\right),
\\
&& M^{(1)}_{D_{12}} =
-\frac{1}{3}(5\langle Q\rangle^{(0)}+8\langle \tilde{Q}_S\rangle^{(0)})
 C_2^bC_8^b\left(1+\epsilon\left(\frac{2}{3}
+ 2\log \frac{\mu_1}{m_b}\right)\right).
\end{eqnarray}

\subsection{Penguin operators}
For the matrix elements with two QCD penguin operators we write
\begin{equation}
  M_{\rm peng} = -\frac{G^2_F m_b^2}{12\pi} (V^*_{cs}V_{cb})^2 \,
\lt[ \sum_{j=3}^6 M_{j2} +
\sum_{\stackrel{j=3}{\scriptscriptstyle j\leq k}}^6 M_{jk} \rt]
\end{equation}
As usual we expand $M_{jk}$ as
$M_{jk}=M_{jk}^{(0)}+\frac{\alpha_s}{4\pi} M_{jk}^{(1)}+\ldots$.
The unrenormalized LO and NLO matrix elements necessary for the
renormalization of the penguin diagrams $D_{11}$ and $D_{12}$ are the
following:
\begin{eqnarray}
&& M_{32}^{(0)} = 2C_2^bC_3^b F_3,\qquad M_{42}^{(0)} =
 2C_2^bC_4^b F_4, \nn && M_{52}^{(0)} = 0,
\qquad M_{62}^{(0)} = 0, \label{eq:m2i} \\
&&M_{42}^{(1)} =  \,
 2C_2^bC_4^b (F_1+F_2), \qquad M_{62}^{(1)} =
 \, 2C_2^bC_6^b (F_1+F_2) \label{eq:m2426}
\end{eqnarray}
where
\begin{eqnarray}
&& F_1 = -\frac{1}{9}(8 \langle \tilde{Q}_S\rangle^{(0)} + 5 \langle Q\rangle^{(0)}) \left[\frac{1}{2 \epsilon
} + 2 \log \frac{\mu_1}{m_b} + \frac{1}{6} \left(19-3 \sqrt{3} \pi
\right)\right.
\\
&& \hspace{2.5cm} \nonumber + \epsilon  \left(\frac{1}{4} \sqrt{3}
\pi  \log 3 - \frac{\pi ^2}{12} + \left(\frac{19}{3}-\sqrt{3} \pi
\right) \left(2 \log \frac{\mu_1}{m_b} + \frac{3}{2}\right) + 4 \log
^2 \frac{\mu_1}{m_b}\right.
\\
&& \left.\left. \nonumber \hspace{3.5cm} - \frac{3}{2} i \sqrt{3}
\left(\text{Li}_2\left(\frac{1}{2}-\frac{i
}{2\sqrt{3}}\right)-\text{Li}_2\left(\frac{1}{2}+\frac{i
}{2\sqrt{3}} \right)\right)\right)\right],
\\
&& F_2 = -\frac{1}{9}(8 \langle \tilde{Q}_S\rangle^{(0)} + 5
\langle Q\rangle^{(0)})\left[\frac{\frac{1}{2}+\sqrt{1-4z}
\left(\frac{1}{2}+z\right)}{\epsilon}+2\log\frac{\mu_1}{m_b}+\frac{7}{6}+2z-\frac{1}{2}\log(1-4z)
\right.
\\
&& \nonumber \hspace{1cm} + \log(1-\sigma) - \frac{\log\sigma}{2} +
\frac{1}{6}\sqrt{1-4z}\left(7+20z+3(2z+1)
\left(4\log\frac{\mu_1}{m_b}+\log\sigma-\log(1-4z)\right)\right)
\\
&& \nonumber \hspace{0.8cm} +
\epsilon\frac{1}{12}\left(34-\pi^2+80z-2\log(1-4z)\left(12\log\frac{\mu_1}{m_b}+7+12z+6\log(1-\sigma)
-3\log\sigma\right)\right.
\\
&& \nonumber \hspace{1cm} + 8
\left(7+12z+6\log(1-\sigma)-3\log\sigma\right)\log\frac{\mu_1}{m_b}
+ 48\log ^2\frac{\mu_1}{m_b} +3 \log ^2(1-4z)
\\
&& \nonumber \hspace{1cm} + (2 \log (1-\sigma )-\log\sigma ) (24 z+6
\log (1-\sigma )-3 \log\sigma+14)
\\
&&  \nonumber \hspace{1cm} + \sqrt{1-4z}\left(34+108 z+2 (20 z+7)
\left(4 \log \frac{\mu_1}{m_b}-\log(1-4 z)+\log\sigma\right)\right.
\\
&&  \nonumber \left.\left.\left. +3 (2 z+1) \left(\left(4
\log\frac{\mu_1}{m_b}-\log (1-4 z)+\log \sigma \right)^2-4
\text{Li}_2 (\sigma)-2 \log ^2 \sigma -3 \pi
^2\right)\right)\right)\right],
\end{eqnarray}
and 
\begin{eqnarray}
\hspace{-8mm}
 F_3 \!\!&=&\!\! \left(\frac{1}{2} \langle Q\rangle^{(0)} - \langle \tilde{Q}_S\rangle^{(0)}\right) \left[1+\epsilon
\left(\frac{2}{3}+2\log \frac{\mu_1}{m_b}\right)+\epsilon ^2 \left(2
\log ^2 \frac{\mu_1}{m_b}+\frac{4}{3} \log
\frac{\mu_1}{m_b}-\frac{\pi^2}{4}+\frac{13}{9}\right)\right]\! ,
\\
\hspace{-8mm}
F_4 \!\!&=&\!\! \left(\langle Q\rangle^{(0)} + \langle \tilde{Q}_S\rangle^{(0)}\right)
 \left[1+\epsilon  \left(\frac{2}{3}+2\log \frac{\mu_1}{m_b}\right)+\epsilon ^2 \left(2 \log ^2 \frac{\mu_1}{m_b}+\frac{4}{3} \log \frac{\mu_1}{m_b}-\frac{\pi^2}{4}+\frac{13}{9}\right)\right].
\end{eqnarray}
We only need the LO contributions $M_{jk}^{(0)}$ for $j,k\geq 3:$
\begin{eqnarray}
\label{QCDpeng}  \nonumber M^{(0)}_{33} =
 3 C_3^{b2}
\hat F_5, &&\qquad M^{(0)}_{34} = 2C_3^bC_4^b \hat F_5,
\\
 \nonumber M^{(0)}_{35} =
6 C_3^bC_5^b \hat F_7,&&\qquad M^{(0)}_{36} = 2C_3^bC_6^b \hat F_7,
\\
 M^{(0)}_{44} =
C_4^{b2} \hat F_6,&&\qquad M^{(0)}_{45} = 2C_4^bC_5^b \hat F_7,
\\
 \nonumber M^{(0)}_{46} =
2C_4^bC_6^b \hat F_7,&&\qquad M^{(0)}_{55} = 3 C_5^{b2} \hat F_5,
\\
 \nonumber M^{(0)}_{56} =
2C_5^bC_6^b \hat F_5,&&\qquad M^{(0)}_{66} = C_6^{b2} \hat F_6,
\end{eqnarray}
We need the coefficient functions  up to
$\epsilon^2$, finding
\begin{eqnarray}
&& \nonumber \hat F_5 = \sqrt{1-4z}\left[\frac{1}{2} \langle Q\rangle^{(0)}(1-4z) -
\langle \tilde{Q}_S\rangle^{(0)}(2z+1)\right.
\\
&& \nonumber  \hspace{2.5cm} + \frac{1}{3}\epsilon\left(\frac{1}{2} \langle Q\rangle^{(0)}
(1-4z)\left(5+6\log
\frac{\mu_1}{m_b}-3\log(1-4z)\right)\right.
\\
&& \left. \nonumber \hspace{4cm} - \langle \tilde{Q}_S\rangle^{(0)}
\left(5+16z-3(2z+1)\left(2\log \frac{\mu_1}{m_b}-\log(1-4
z)\right)\right)\right)
\\
&& \nonumber + \frac{1}{36}\epsilon^2\left(\frac{1}{2} \langle Q\rangle^{(0)}
(1-4z)\left(112-9\pi^2+60\left(2\log
\frac{\mu_1}{m_b}-\log(1-4z)\right)\right.\right.
\\
&& \nonumber \left.\left. \hspace{4cm} +18\left(2\log
\frac{\mu_1}{m_b}-\log(1-4z)\right)^2\right)\right.
\\
&& \nonumber \hspace{1.5cm} - \langle \tilde{Q}_S\rangle^{(0)}
\left(112+416z-9\pi^2(2z+1)+12(16z+5)\left(2\log
\frac{\mu_1}{m_b}-\log(1-4 z)\right)\right.
\\
&& \left.\left.\left. \hspace{3cm} +18(2z+1)\left(2\log
\frac{\mu_1}{m_b}-\log(1-4z)\right)^2\right)\right)\right],
\\
&& \nonumber \hat F_6 = \sqrt{1-4z}\left[\langle Q\rangle^{(0)}(1-z) + \langle \tilde{Q}_S\rangle^{(0)}(2z+1)
\right.
\\
&& \nonumber  \hspace{2.5cm} +\frac{1}{3}\epsilon\left(\langle Q\rangle^{(0)} \left(5-2z + 3(1-z)(2\log
\frac{\mu_1}{m_b}-\log(1-4z))\right)\right.
\\
&& \left. \nonumber \hspace{4cm} + \langle \tilde{Q}_S\rangle^{(0)}
\left(5+16z+3(2z+1)\left(2\log \frac{\mu_1}{m_b}-\log(1-4
z)\right)\right)\right)
\\
&& \nonumber \hspace{1cm} + \frac{1}{36}\epsilon^2\left(\langle Q\rangle^{(0)}
\left(112-16z-9\pi^2(1-z)+12(5-2z)\left(2\log
\frac{\mu_1}{m_b}-\log(1-4z)\right)\right.\right.
\\
&& \left. \nonumber \hspace{3cm} +18(1-z)\left(2\log
\frac{\mu_1}{m_b}-\log(1-4z)\right)^2\right)
\\
&& \nonumber \hspace{2.0cm} + \langle \tilde{Q}_S\rangle^{(0)}
\left(112+416z-9\pi^2(2z+1) +12(16z+5)\left(2\log
\frac{\mu_1}{m_b}-\log(1-4 z)\right)\right.
\\
&& \left.\left.\left. \hspace{3cm} +18(2z+1)\left(2\log
\frac{\mu_1}{m_b}-\log(1-4z)\right)^2\right)\right)\right],
\\
&& \nonumber \hat F_7 = \langle Q\rangle^{(0)} z\sqrt{1-4z}\left(3+3\epsilon\left(2+2\log
\frac{\mu_1}{m_b}-\log(1-4z)\right)\right.
\\
&& \left. \hspace{1cm} +
\frac{3}{4}\epsilon^2\left(\pi^2-16+2\left(4+2\log
\frac{\mu_1}{m_b}-\log(1-4z)\right)\left(2\log
\frac{\mu_1}{m_b}-\log(1-4z)\right)\right)\right).
\end{eqnarray}
The $M_{25}^{(0)} $ and $M_{26}^{(0)}$ (due to the
$(V-A)\otimes (V+A)$ chiral structure) are proportional to $m_c^2$
and vanish in our approximation $m_c=0$ for the charm
lines attached to weak vertices.

\section{Results of master integrals}
We have reduced the Feynman diagrams shown on Fig.\ref{fig1} to master
integrals by means of the program FIRE \cite{Smirnov:2008iw}.  For the
full-theory diagrams we have calculated the absorptive part of master
integrals, i.e. the 2-, 3-, 4- particle cuts with a massive $c$-,
$b$-quarks in the closed fermion loop, with a massive gluon in infrared
singular diagrams and a massless $c$-quark in the weak loop, using
formulas for phase space integrals derived in \cite{Asatrian:2012tp}.
For some integrals, with massive charm, we used a Mellin-Barnes
representation \cite{Boos:1990rg} and expanded in terms of the small
parameter $z=m_c^2/m_b^2$. The master integrals, which include $m_g$, are expanded
over $z_g=m_g^2/m_b^2$. The results of master integrals have been
checked numerically by means of the program SecDec-3
\cite{Borowka:2015mxa}.

The results for diagrams with massless $u$-, $d$-, $s$-quarks in the
closed fermion loop are obtained by taking the limit $m_c\to 0$ in the
results with a massive $c$-quark in the closed fermion loop.

From the results below one can see that the first three orders in the
expansion over $z$ already exhibit a good convergence.

Our convention for the loop measure is
\begin{eqnarray}
\int [dk]=\int \frac{dk_1^d}{(2\pi)^d}\int
\frac{dk_2^d}{(2\pi)^d}\int \frac{dk_3^d}{(2\pi)^d} .
\end{eqnarray}

Some of the integrals in the following subsections have more than one cut
(e.g. 2, 3 and 4 particle cuts). The following subsections quote the
results of the various cuts. We write
\begin{displaymath}
 \imag = \imag^{(2)}+\imag^{(3)}+\imag^{(4)}
\end{displaymath}
to separate the contributions from these cuts.

\subsection{Results for the four-particle cuts of the master integrals}

\begin{eqnarray}
\nonumber && \hspace{-1.5cm} \imag^{(4)} \int [dk] \frac{1}{\left(k_2^2-m_b^2\right)
k_3^2\left((k_1-p_b)^2-m_c^2\right)\left((k_1-k_2)^2-m_c^2\right)(k_2-k_3)^2}
\\
\nonumber &=& \frac{m_b^{2-6\epsilon}}{8192\pi^5}\left[ \frac{2z^7(35\log
z+611)}{1225}+\frac{1}{200}z^6(20\log z+151)+z^5\left(\frac{\log
z}{5}+\frac{559}{900}\right) \right.
\\
&& \nonumber \hspace{0cm} +z^4\left(\frac{\log
z}{2}+\frac{7}{24}\right) +z^3\left(2\log
z-\frac{11}{3}\right)+z^2\left(\log^2 z-7\log z+\frac{27}{2}\right)
\\
&& \nonumber -\frac{2}{3}z\left(6\log
z+\pi^2+6\right)+\frac{\pi^2}{3}-\frac{7}{2}
\\
&& \nonumber \hspace{-0.5cm} + \epsilon \left(z^7 \left(-0.0792168\log
z-1.3829\right) +z^6\left(-0.138629\log z-1.04665\right)\right.
\\
&& \nonumber \hspace{0cm} +z^5\left(-0.277259\log z-0.861043\right)
+z^4\left(4.3\log z-8.65202\right) +z^3\left(-1.33333\log z-14.4059\right)
\\
&& \nonumber \hspace{0cm}
+z^2\left(-\log^3 z+4.5\log^2 z-11.6595\log z+33.472\right) +z\left(2\log^2 z-32\log z-102.311\right)
\\
&& \left.\left. \hspace{0cm}
-3.00788z^{7/2}-21.0552z^{5/2}+105.276z^{3/2}-2.50034\right) \right]
+ \mathcal{O}\left(z^8,\epsilon^2\right),
\end{eqnarray}
\begin{eqnarray}
\nonumber && \hspace{-1.5cm}  \imag^{(4)} \int [dk] \frac{1}{\left(k_2^2-m_g^2\right)
(k_3^2-m_c^2) (k_1-p_b)^2(k_1-k_2)^2\left((k_2-k_3)^2-m_c^2\right)}
\\
&=& \nonumber \frac{m_b^{2-6\epsilon}}{8192\pi^5}\left[-\frac{2\pi^2
z}{3}+\frac{1}{2}\sqrt{1-4z}(2z-9)
+2z\log^2\sigma-8z\text{Li}_2(-\sigma) \right.
\\
&& \nonumber \hspace{1cm} +\log\sigma\left(-1-4z-2
z^2-8z\log(1-\sigma)+4z\log(1-4z)\right)
\\
&& \nonumber \hspace{-0.5cm}
+z_g\left(4\text{Li}_2(-\sigma)-2\log\sigma(z+2+\log(1-4z)-2\log(1-\sigma))+\frac{\sqrt{1-4
z}(1-58z)}{6z}+\frac{\pi^2}{3}-\log^2\sigma\right)
\\
&& \nonumber \hspace{-0.5cm} +z_g^2\left(\frac{\sqrt{1-4
z}\left(128z^2+166z+3\right)}{180z^2}+\left(\frac{1}{3z}+1\right)\log\sigma\right)
\\
&& \left. \hspace{-0.5cm} +z_g^3\left(\frac{\sqrt{1-4z}(384 z^3-512
z^2+464 z+15)}{6300z^3}+\frac{\log\sigma}{30z^2}\right)\right] +
\mathcal{O}\left(z_g^4,\epsilon^1\right),
\end{eqnarray}
\begin{eqnarray}
\nonumber && \hspace{-1.5cm} \imag^{(4)} \int [dk] \frac{1}{(k_3^2-m_c^2)
(k_1-p_b)^2(k_1-k_2)^2\left((k_2-k_3)^2-m_c^2\right)}
\\
&=& \nonumber \frac{m_b^{4-6\epsilon}}{49152\pi^5}\left[-48z^2\text{Li}_2(-\sigma)+12z^2\log\sigma(2\log(1-4z)-4\log(1-\sigma)+\log\sigma)
\right.
\\
&& \left. \hspace{1.5cm} -4\pi^2
z^2+\sqrt{1-4z}\left(12z^2+20z+1\right)+12(1-z)(2z+1)z\log\sigma\right]
+ \mathcal{O}\left(\epsilon^1\right),
\end{eqnarray}
\begin{eqnarray}
\nonumber && \hspace{-1.5cm} \imag^{(4)} \int [dk] \frac{1}{\left(k_2^2-m_g^2\right)
(k_3^2-m_c^2) (k_1-p_b)^2\left((k_2-p_b)^2-m_b^2\right)(k_1-k_2)^2\left((k_2-k_3)^2-m_c^2\right)}
\\
&=& \nonumber \frac{m_b^{-6\epsilon}}{8192\pi^5}\left[z^6\left(\frac{\log z}{55}+\frac{958}{3025}\right)
+z^5\left(\frac{\log z}{30}+\frac{1349}{5400}\right)
+z^4\left(\frac{\log z}{14}+\frac{761}{3528} \right)\right.
\\
&&  \nonumber \hspace{-1cm}+z^3 \left(\frac{\log z}{5}+\frac{13}{150}\right) +z^2\left(\log
z-\frac{13}{6}\right)+z\left(\log^2 z-10\log
z+30\right)-4\pi^2 \sqrt{z}+2\log z+\frac{\pi^2}{3}+8
\\
&& \nonumber \hspace{-1.5cm} +z_g\left(z^6\left(\frac{\log z}{286}+\frac{90943}{613470}\right)
+z^5\left(\frac{\log z}{165}+\frac{17189}{163350}\right)
+z^4\left(\frac{\log z}{84}+\frac{1867}{21168}\right)
+z^3 \left(\frac{\log z}{35}+\frac{904}{11025}\right)
\right.
\\
&& \nonumber \left.\hspace{0cm} +\frac{1}{100}z^2(10\log z+1)+z\left(\log z-\frac{11}{3}\right) +\frac{\pi^2
\sqrt{z}}{2}-\frac{\log^2 z}{2}+2\log z-6+\frac{\pi^2}{2\sqrt{z}}-\frac{1}{3z}\right)
\\
&& \nonumber \hspace{-1.5cm}
+z_g^2\left(\frac{\pi^2}{32z^{3/2}}+z^6\left(\frac{\log
z}{1430}+\frac{3319103}{42942900}\right)+z^5\left(\frac{\log
z}{858}+\frac{544943}{11042460}\right)+z^4\left(\frac{\log
z}{462}+\frac{119729}{3201660}\right)\right.
\\
&& \nonumber \hspace{0cm} +z^3\left(\frac{\log z}{210}+\frac{4573}{132300}\right)
+z^2\left(\frac{\log z}{70}+\frac{799}{22050}\right)+\frac{z}{50}(5\log z-7)+\frac{\pi^2 \sqrt{z}}{32}
+ \frac{1}{6}\left(2-3\log z\right)
\\
&& \left.\left. \hspace{0cm} -\frac{\pi^2}{16\sqrt{z}}+\frac{3\log
z-4}{18z}-\frac{1}{30 z^2}\right)\right] +
\mathcal{O}\left(z^7,z_g^3,\epsilon^1\right),
\end{eqnarray}
\begin{eqnarray}
\nonumber && \hspace{-1.5cm} \imag^{(4)} \int [dk] \frac{1}{k_1^2 (k_1-p_b)^2
\left((k_2-p_b)^2-m_b^2\right)(k_1-k_2)^2\left(k_3^2-m_c^2\right)\left((k_2-k_3)^2-m_c^2\right)}
\\
&=& \nonumber \frac{m_b^{-6\epsilon}}{8192\pi^5}\left[z^5(0.075\log z+0.385)
+z^4(0.166667\log z+0.310185) +z^3(0.5\log z-0.166667)\right.
\\
&& \nonumber \hspace{0.3cm} \left. +z^2(3\log z-9.5)
+z\left(-0.333333\log^3 z+\log^2 z+4.57974\log
z+9.84495\right)-0.710132\right]
\\
&& + \mathcal{O}\left(z^6,\epsilon^1\right),
\end{eqnarray}
\begin{eqnarray}
\nonumber && \hspace{-1.5cm} \imag^{(4)} \int [dk] \frac{1}{k_1^2 k_2^2 (k_1-p_b)^2
(k_2-p_b)^2\left((k_1-k_3)^2-m_c^2\right)\left((k_2-k_3)^2-m_c^2\right)}
\\
&=& \nonumber \frac{m_b^{-6\epsilon}}{2048\pi^5}\left[z^7\left(\log
z\left(44\log z+\frac{27634}{315}\right)-22\pi
^2-\frac{209379}{1225}\right)
+\frac{1}{6}\left(\pi^2-6\right)\right.
\\
&& \nonumber \hspace{-1cm} +z^6\left(-\log z\left(\frac{84\log
z}{5}+\frac{466}{15}\right)+\frac{42\pi^2}{5}+\frac{75343}{1125}\right)
+z^5\left(\log
z\left(7\log(z)+\frac{172}{15}\right)-\frac{7\pi^2}{2}-\frac{104551}{3600}\right)
\\
&& \nonumber \hspace{-1cm} +z^4 \left(-\log z\left(\frac{10\log
z}{3}+\frac{13}{3}\right)+\frac{5\pi
^2}{3}+\frac{511}{36}\right)+z^3\left(\log z\left(2\log
z+\frac{4}{3}\right)-\pi ^2-\frac{161}{18}\right)
\\
&& \left. \nonumber \hspace{-1cm} +z^2\left(2(1-\log z)\log
z+\pi^2+5\right)+\frac{1}{3} z\left(\log z\left((\log z-3)\log
z-2\pi^2+6\right)-30\zeta(3)+2\pi ^2-6\right) \right]
\\
&& + \mathcal{O}\left(z^8,\epsilon^1\right),
\end{eqnarray}
\begin{eqnarray}
\nonumber && \hspace{-1.5cm} \imag^{(4)} \int [dk] \frac{(k_3.p_b) }{(k_3^2-m_c^2)
(k_1-p_b)^2(k_1-k_2)^2\left((k_2-k_3)^2-m_c^2\right)}
\\
&=& \nonumber \frac{m_b^{6-6\epsilon}}{49152\pi^5}\left[2z\left(-5z^3+6z^2-3z+1\right)\log\sigma
+ \frac{1}{12}\sqrt{1-4z}\left(60z^3-62z^2+26z+3\right)\right.
\\
&& \nonumber + \epsilon \left(\frac{71}{24}+z^8 \left(-\frac{858 \log z}{35}-\frac{570523}{7350}\right)+z^7 \left(-\frac{88 \log z}{5}-\frac{70991}{1575}\right) + z^6 \left(-\frac{84 \log z}{5}-\frac{8177}{300}\right)\right.
\\
&& \nonumber \hspace{0.5cm} +z^5 \left(\frac{541}{60}-28 \log z\right) +z^4 \left(-\frac{67 \log z}{3}+\frac{35 \pi^2}{3}+\frac{737}{12}\right) +z^3 \left(62 \log z-14 \pi ^2+\frac{91}{3}\right)
\\
&& \left.\left. \nonumber \hspace{0.4cm} + z^2 \left(3 \log ^2 z-36 \log z+6 \pi ^2-\frac{105}{2}\right) +z \left(-\log^2 z+15 \log z-2 \pi ^2+\frac{427}{18}\right)\right)\right]
\\
&& +\mathcal{O}\left(z^9,\epsilon^2\right),
\end{eqnarray}
\begin{eqnarray}
&& \nonumber \hspace{-1.5cm} \imag^{(4)} \int [dk]
\frac{k_1^2}{k_2^2(k_1-p_b)^2\left((k_1-k_3)^2-m_c^2\right)\left((k_2-k_3)^2-m_c^2\right)}
\\
&=& \nonumber \frac{m_b^{6-6\epsilon}}{294912\pi^5}
\left(24z\left(-5z^3+6z^2-3z+1\right)\log\sigma+\sqrt{1-4z}\left(60z^3-62
z^2+26z+3\right)\right)
\\
&& +\mathcal{O}\left(\epsilon^1\right).
\end{eqnarray}

\subsection{Results for the three-particle cuts of the master integrals}
\begin{eqnarray}
\nonumber &&\hspace{-1.5cm}\imag^{(3)}\int [dk] \frac{1}{(p_b-k_1)^2
(k_1-k_2)^2 k_2^2 (k_3^2-m_c^2)}
\\
&=& \nonumber
\frac{m_b^{4-6\epsilon}z}{8192\pi^5}\left[-\frac{1}{\epsilon}+\log
z-\frac{15}{2} +\frac{1}{4}\epsilon\left(30 \log z-2\log ^2
z+3\pi^2-145\right) \right.
\\
&& \left. \hspace{-0.5cm} +\frac{1}{24}\epsilon^2\left(
45\left(3\pi^2-77\right)+2\log z\left(\log z(2\log
z-45)-9\pi^2+435\right)+264\zeta(3)\right)\right] +\mathcal{O}(\epsilon^3),
\\
\nonumber &&\hspace{-1.5cm}\imag^{(3)}\int [dk] \frac{1}{(p_b-k_1)^2
(k_1-k_2)^2 (k_2^2-m_g^2) (k_3^2-m_c^2)}
\\
&=& \nonumber
\frac{m_b^{4-6\epsilon}z}{8192\pi^5}\left[\frac{z_g^2-2z_g\log
z_g-1}{\epsilon}+\log z-\frac{15}{2}\right.
\\
&& \nonumber \hspace{2cm} + z_g\left(2\log z_g\left(\log
z-4\right)+\log^2 z_g+\frac{4}{3}\left(\pi^2-3\right)\right)
\\
&& \left. \hspace{2cm} + z_g^2\left(-\log z+\log
z_g-\frac{5}{2}\right)+\frac{2z_g^3}{3}\right]
+\mathcal{O}(z_g^4,\epsilon^1),
\\
\nonumber &&\hspace{-1.5cm}\imag^{(3)}\int [dk] \frac{1}{(p_b-k_1)^2
(k_1-k_2)^2 (k_2^2-m_g^2) (k_3^2-m_c^2)
\left((p_b-k_2)^2-m_b^2\right)}
\\
&=& \nonumber
\frac{m_b^{2-6\epsilon}z}{4096\pi^5}
\left[\frac{-2\pi\sqrt{z_g}+z_g(4-\log z_g)+2}{2\epsilon} \right.
\\
&& \nonumber \hspace{2cm} -\log z+8 +\pi\sqrt{z_g} \left(\log
z+\log(4z_g)-5\right)
\\
&&\left. \hspace{2cm} +\frac{1}{12}z_g\left(6(\log
z_g-4)\log z+3(\log z_g-1)^2+4\pi^2-15\right) \right]
+\mathcal{O}(z_g^{3/2},\epsilon^1),
\\
\nonumber &&\hspace{-1.5cm}\imag^{(3)}\int [dk] \frac{1}{(p_b-k_1)^2
(k_1-k_2)^2 (k_2^2-m_g^2) (k_3^2-m_c^2)
\left((k_2-k_3)^2-m_c^2\right)}
\\
&=& \nonumber
\frac{m_b^{2-6\epsilon}}{8192\pi^5}\left[\frac{z_g^2-2z_g\log
z_g-1}{\epsilon}-\frac{13}{2}+\log z \right.
\\
&& \nonumber \hspace{2cm} +z_g\left(2\log z_g\left(\log
z-3\right)-\frac{1}{6z}+\log^2
z_g+\frac{4}{3}\left(\pi^2-3\right)\right)
\\
&&\left.  \hspace{2cm} +z_g^2\left(-\frac{(1-3z)\log
z_g}{3z}-\log z-\frac{1}{60z^2}-\frac{7}{2}\right) \right]
+\mathcal{O}(z_g^{3},\epsilon^1),
\\
\nonumber &&\hspace{-1.5cm}\imag^{(3)}\int [dk] \frac{1}{\left(k_2^2-m_g^2\right)
(k_3^2-m_c^2) (k_1-p_b)^2\left((k_2-p_b)^2-m_b^2\right)(k_1-k_2)^2\left((k_2-k_3)^2-m_c^2\right)}
\\
&=& \nonumber
\frac{m_b^{-6\epsilon}}{4096\pi^5}\left[\frac{-2\pi\sqrt{z_g}+z_g(4-\log
z_g)+2}{2\epsilon}\right.
\\
&& \nonumber \hspace{2cm} -\log z+7 + \pi \sqrt{z_g}\left(\log
z+\log(4 z_g)-4\right)
\\
&&\left.  \hspace{2cm} + \frac{1}{12}z_g\left(6 (\log
z_g-4)\log z+3\log^2 z_g+4\pi^2-36+\frac{2}{z}\right) \right]
+\mathcal{O}(z_g^{3/2},\epsilon^1),
\end{eqnarray}

\subsection{Results for the two-particle cuts of the master integrals}
\begin{eqnarray}
\nonumber
&& \imag^{(2)} \int [dk] \frac{1}{{k_1^2 k_2^2 (k_1 - p_b)^2 (k_2 - p_b)^2 \left((k_1 - k_3)^2 - m_c^2\right) \left((k_2 - k_3)^2 -m_c^2\right)}}
\\
&&\nonumber
=\frac{m_b^{-6\epsilon}}{8192\pi^5}
\left[\frac{2}{\epsilon^2}+\frac{14}{\epsilon}-z^8\left(\frac{1144}{7}\pi^2+\frac{324314461}{617400}-
\frac{54031}{105}\log z -\frac{1716}{7}\log^2 z\right) \right.
\\
&& \nonumber
+z^7\left(\frac{176\pi^2}{3}+\frac{2168531}{11025}-\frac{55268}{315}\log z-88\log^2 z\right) -z^6\left(\frac{112\pi^2}{5}+\frac{88799}{1125}-\frac{932}{15}\log z-\frac{168}{5}\log^2 z\right)
\\
&& \nonumber +z^5\left(\frac{28\pi^2}{3}+\frac{31363}{900}-\frac{344}{15}\log z-14\log^2 z\right)-z^4\left(\frac{40\pi^2}{9}+\frac{953}{54}-\frac{26}{3}\log z -\frac{20}{3}\log^2 z\right)
\\
&& \nonumber +z^3\left(\frac{8\pi^2}{3}+\frac{98}{9} -\frac{8}{3}\log z -4\log^2 z\right) -z^2\left(\frac{8\pi^2}{3}+6 +4\log z -4\log^2 z\right)
\\
&&  \left. + z\left(16\zeta(3)-\frac{8\pi^2}{3}+8 +8\left(\frac{\pi ^2}{3}-1\right)\log z +4\log^2 z -\frac{4}{3}\log^3 z\right) -\frac{25\pi^2}{6}+66 \right] +\mathcal{O}(z^{9},\epsilon^1),
\end{eqnarray}
\begin{eqnarray}
\nonumber
&& \hspace{-1.5cm} \imag^{(2)} \int [dk] \frac{1}{{k_1^2 (k_1 - p_b)^2 (k_3^2 - m_c^2) \left((k_2 - p_b)^2 - m_b^2\right) \left((k_2 - k_3)^2-m_c^2\right)}}
\\
&=& \nonumber
\frac{m_b^{2-6\epsilon}}{8192\pi^5}\left[\frac{-2z-1}{\epsilon^2}
+ \frac{2z\left(2\log z-5\right)-\frac{9}{2}}{\epsilon}
+z^8\left(\frac{73}{14112}-\frac{\log z}{84}\right) \right.
\\
&& \nonumber +z^7\left(\frac{107}{11025}-\frac{2}{105}\log z\right)
+z^6\left(\frac{37}{1800}-\frac{\log z}{30}\right)
+z^5\left(\frac{47}{900}-\frac{\log z}{15}\right)
\\
&& \nonumber +z^4\left(\frac{13}{72}-\frac{\log z}{6}\right) +z^3\left(\frac{11}{9}-\frac{2}{3}\log z\right)+z^2 \left(-\frac{2\pi^2}{3}-\frac{7}{2}+3\log
z -\log^2 z\right)
\\
&& \left. +z\left(\frac{3\pi^2}{2}-30+20\log
z-2\log^2 z\right) - \frac{7\pi^2}{12}-\frac{47}{4}\right] +\mathcal{O}(z^{9},\epsilon^1),
\end{eqnarray}
\begin{eqnarray}
\nonumber
&& \hspace{-1.5cm} \imag^{(2)} \int [dk] \frac{1}{k_1^2 (k_1 - p)^2 (k_2^2 - m_g^2)
(k_3^2 - m_c^2) \left((k_2 - p_b)^2 - m_b^2\right) \left((k_2 - k_3)^2 - m_c^2\right)}
\\
&=& \nonumber
\frac{m_b^{-6\epsilon}}{8192\pi^5}\left[-\frac{1}{\epsilon^2}+\frac{2\pi\sqrt{z_g}+z_g(\log
z_g-2)-7}{\epsilon } +z^6\left(\frac{181}{54450}-\frac{\log
z}{165}\right) \right.
\\
&& \nonumber +z^5\left(\frac{121}{16200}-\frac{\log z}{90}\right) +z^4\left(\frac{73}{3528}-\frac{\log z}{42}\right)+z^3\left(\frac{37}{450}-\frac{\log z}{15}\right) + z^2\left(\frac{13}{18}-\frac{1}{3}\log z\right)
\\
&& \left. +4\pi^2\sqrt{z}+z\left(-\frac{2\pi^2}{3}-14+6\log z-\log^2 z\right)-\frac{7\pi^2}{12}-33\right] +\mathcal{O}(z^{7},\epsilon^1),
\end{eqnarray}
\begin{eqnarray}
\nonumber
&& \hspace{-1.5cm} \imag^{(2)} \int [dk] \frac{1}{k_1^2 (k_1 - p_b)^2 (k_2^2 - m_g^2) (k_3^2 - m_c^2) \left((k_2 - k_3)^2 - m_c^2\right)}
\\
&=& \nonumber
\frac{m_b^{2-6\epsilon}}{8192\pi^5}\left[\frac{-2z-z_g}{\epsilon^2}+\frac{2z(2\log
z-5)+z_g(2\log z_g-5)}{\epsilon}\right.
\\
&& \nonumber +\frac{1}{6}z \left(-24\log^2 z+120\log
z+\pi^2-204\right)
\\
&& \left. +z_g\left(-2\log z(\log z_g-2)+\log^2 z-\log^2 z_g+6\log
z_g+\frac{\pi^2}{12}-9\right)\right]+\mathcal{O}(z_g^{2},\epsilon^1),
\end{eqnarray}
\begin{eqnarray}
\nonumber && \hspace{-1.5cm} \imag^{(2)} \int [dk] \frac{1}{k_1^2 (k_1-p_b)^2 (k_2^2 - m_g^2) \left((k_2 - p_b)^2 - m_b^2\right) (k_3^2-m_c^2)}
\\
\nonumber &=&\frac{m_b^{2-6\epsilon}
z}{8192\pi^5}\left[-\frac{2}{\epsilon^2}+\frac{2(\log
z-5)-\frac{1}{4}\pi
z_g^{3/2}+\frac{z_g^2}{6}+2\pi\sqrt{z_g}+z_g(\log
z_g-2)}{\epsilon}\right.
\\
&& \nonumber -\log^2 z+10\log z+\frac{\pi^2}{6}-34-2\pi
\sqrt{z_g}(\log z+\log(4z_g)-5)
\\
&& \nonumber +\frac{1}{2}z_g\left(-2\log z\log z_g+4\log z-\log^2
z_g+6\log z_g-4\right)+\frac{1}{4}\pi z_g^{3/2}\left(\log
z+\log(4z_g)-3\right)
\\
&& \left. +\frac{1}{6}z_g^2\left(-\log z-3\log(4z_g)+19-\log
4\right)\right]+\mathcal{O}(z_g^{5/2},\epsilon^1),
\end{eqnarray}
\begin{eqnarray}
\nonumber && \hspace{-1.4cm} \imag^{(2)} \int [dk] \frac{1}{k_1^2 (k_1-p_b)^2(k_3^2-m_c^2)\left((k_2-p_b)^2-m_b^2\right)\left((k_2-k_3)^2-m_c^2\right)}
\\
\nonumber &=&
\frac{m_b^{2-6\epsilon}}{4096\pi^5}\left[\frac{z+\frac{1}{2}}{\epsilon
^2}+\frac{z(5-2\log z)+\frac{9}{4}}{\epsilon}+z^8\left(\frac{\log
z}{168}-\frac{1}{28224}73\right)+z^7\left(\frac{\log
z}{105}-\frac{107}{22050}\right)\right.
\\
&& \nonumber +z^6\left(\frac{\log z}{60}-\frac{37}{3600}\right)+z^5\left(\frac{\log
z}{30}-\frac{47}{1800}\right)+z^4\left(\frac{\log z}{12}-\frac{13}{144}\right)+z^3\left(\frac{\log z}{3}-\frac{11}{18}\right)
\\
&& \nonumber +z^2\left(\frac{\log^2 z}{2}-\frac{3}{2}\log z+\frac{\pi^2}{3}+\frac{7}{4}\right)+z\left(\log^2 z-10\log
z-\frac{3\pi^2}{4}+15\right)+\frac{7\pi^2}{24}+\frac{47}{8}
\\
&& \nonumber \hspace{-1cm}
+\epsilon\left(-\frac{16}{315}\pi^2z^{9/2}-\frac{16}{105}\pi^2z^{7/2}-\frac{16}{15}\pi^2
z^{5/2}+\frac{16}{3}\pi^2z^{3/2}-\frac{16\pi^2z^{15/2}}{2145}-\frac{16\pi^2z^{13/2}}{1287}-\frac{16}{693}\pi^2
z^{11/2}\right.
\\
&& \nonumber +z^8\left(\frac{890041\log
z}{30270240}+\frac{\pi^2}{168}-\frac{113307356143}{10908183686400}\right)+z^7
\left(\frac{62281\log
z}{1455300}+\frac{\pi^2}{105}-\frac{344223461}{20170458000}\right)
\\
&& \nonumber +z^6\left(\frac{2473\log
z}{37800}+\frac{\pi^2}{60}-\frac{1403863}{47628000}\right)
+z^5\left(\frac{661}{6300}\log z+\frac{\pi^2}{30}-\frac{44969}{882000}\right)
\\
&& \nonumber +z^4\left(\frac{29}{180}\log z+\frac{\pi^2}{12}-\frac{49}{1200}\right) +z^3\left(\frac{\pi^2}{3}+\frac{17}{9}-\frac{2}{9}\log z\right)
\\
&& \nonumber +\frac{1}{8}z^2\left(-4\log^3 z+18\log^2 z-10\log z+32\zeta(3)-67\right)
\\
&& \nonumber +z \left(-\frac{1}{3}\log^3 z+5\log^2
z+\left(\frac{\pi^2}{6}-34\right)\log z-11\zeta(3)-\frac{15\pi^2}{4}+33\right)
\\
&& \left.\left.
+\frac{5}{2}\zeta(3)+\frac{133}{16}+\frac{21\pi^2}{16}\right)\right] +\mathcal{O}(z^{9},\epsilon^2),
\end{eqnarray}
\begin{eqnarray}
\nonumber && \hspace{-1.5cm} \imag^{(2)} \int [dk] \frac{1}{k_1^2 (k_1 - p_b)^2 k_2^2 (k_2 - p_b)^2 (k_3^2 - m_c^2) \left((k_3 - p_b)^2 - m_c^2\right)}
\\
&& \nonumber = \frac{m_b^{-6\epsilon}}{4096 \pi
^5}\left[\frac{\sqrt{1-4
z}+2}{\epsilon ^2}+\frac{\sqrt{1-4 z} (2 \log\sigma -\log (1-4
z)+6)-2 \log z+12}{\epsilon }\right.
\\
&& \nonumber +\log ^2 z-12 \log z-\frac{3 \pi ^2}{2}+48 +\sqrt{1-4
z}\left(4 \text{Li}_2\left(\frac{1}{2}-\frac{1}{2\sqrt{1-4
z}}\right)-\frac{\log ^2\sigma}{2}+12 \log\sigma\right.
\\
&& \nonumber \left.\left. -\log (1-4 z) (\log\sigma+\log
  z+4)-\log\sigma \log z+\frac{1}{2} \log ^2(1-4 z)+\frac{\log ^2 z}{2}-\frac{7 \pi ^2}{2}+20\right)\right]
\\
&& +\mathcal{O}(\epsilon^1).
\end{eqnarray}

\subsection{Results for integrals with a $b$ quark}
The master integrals with a heavy $b$ quark have only one cut
which contributes to the imaginary part.
\begin{eqnarray}
\nonumber &&\hspace{-3.5cm}\imag\int [dk] \frac{1}{(p_b-k_1)^2
(k_1-k_2)^2 (k_2^2-m_g^2) (k_3^2-m_b^2)
\left((k_2-k_3)^2-m_b^2\right)}
\\
&=& \nonumber \frac{m_b^{2-6\epsilon}}{2(4
\pi)^5}\left[\frac{\frac{{z_g}^2}{4}-\frac{1}{2}{z_g}\log{z_g}-\frac{1}{4}}{\epsilon}+{z_g}^2\left(\frac{\log
{z_g}}{6}-\frac{211}{240}\right)\right.
\\
&& \left.+\frac{1}{24}{z_g}\left(6\log^2{z_g}-36\log{z_g}+8\pi^2-25\right)-\frac{13}{8}\right] +\mathcal{O}\left(z_g^{3},\epsilon^1\right),
\end{eqnarray}
\begin{eqnarray}
\nonumber &&\hspace{-2.5cm}\imag\int [dk] \frac{1}{(p_b-k_1)^2
(k_1-k_2)^2 (k_2^2-m_g^2) (k_3^2-m_b^2) ((k_2-k_3)^2-m_b^2)
\left((p_b-k_2)^2-m_b^2\right)}
\\
&=& \nonumber \frac{m_b^{-6\epsilon}}{2(4
\pi)^5}\left[\frac{-\frac{\pi\sqrt{{z_g}}}{2}+{z_g}\left(1-\frac{\log{z_g}}{4}\right)+\frac{1}{2}}{\epsilon}+\frac{7}{2}\right.
\\
&& \left.+\frac{1}{24}{z_g}\left(3\log
^2{z_g}+4\pi^2-34\right)+\frac{1}{2}\pi\sqrt{{z_g}}(\log{z_g}-4+\log4)\right] +\mathcal{O}\left(z_g^{3/2},\epsilon^1\right),
\end{eqnarray}
\begin{eqnarray}
\nonumber &&\hspace{-2.9cm}\imag\int [dk] \frac{1}{(p_b-k_1)^2 k_1^2
(k_2^2-m_g^2) (k_3^2-m_b^2) \left((k_2-k_3)^2-m_b^2\right)}
\\
&=& \nonumber \frac{m_b^{2-6\epsilon}}{2(4
\pi)^5}\left[\frac{\frac{{z_g}}{4}+\frac{1}{2}}{\epsilon^2}+\frac{\frac{1}{4}{z_g}(5-2\log{z_g})+\frac{5}{2}}{\epsilon}+\frac{1}{24}\left(204-\pi^2\right)\right.
\\
&& \left.+\frac{1}{48}{z_g}\left(12\log^2{z_g}-72\log{z_g}-\pi^2+108\right)\right] +\mathcal{O}\left(z_g^{2},\epsilon^1\right),
\end{eqnarray}
\begin{eqnarray}
\nonumber &&\hspace{-2.9cm}\imag\int [dk] \frac{1}{(p_b-k_1)^2 k_1^2
(k_2^2-m_g^2) (k_3^2-m_b^2) \left((k_2-k_3)^2-m_b^2\right)
\left((p_b-k_2)^2-m_b^2\right)}
\\
&=& \nonumber \frac{m_b^{-6\epsilon}}{2(4
\pi)^5}\left[\frac{1}{4\epsilon^2}+\frac{-\frac{\pi\sqrt{{z_g}}}{2}+\frac{1}{4}{z_g}(2-\log{z_g})+\frac{7}{4}}{\epsilon}-\frac{17\pi^2}{48}+\frac{33}{4}\right.
\\
&& \left.-\pi\sqrt{{z_g}}+{z_g}\left(1-\frac{\log{z_g}}{2}\right)\right] +\mathcal{O}\left(z_g^{3/2},\epsilon^1\right),
\end{eqnarray}
\begin{eqnarray}
&&\hspace{-2.25cm}\imag\int [dk] \frac{1}{(p_b-k_1)^2 k_1^2 (k_3^2-m_b^2)
\left((k_2-k_3)^2-m_b^2\right) \left((p_b-k_2)^2-m_b^2\right)}
\\
&=& \nonumber \frac{m_b^{2-6\epsilon}}{2(4\pi)^5}\left[\frac{3}{4 \epsilon ^2}+\frac{29}{8 \epsilon}+\frac{1}{16} \left(175-\pi ^2\right)+\epsilon\left(\frac{765}{32}-\frac{9}{4} \zeta (3)+\frac{35 \pi ^2}{96} \right)\right] +\mathcal{O}(\epsilon^2),
\end{eqnarray}
\begin{eqnarray}
&&\hspace{-4.1cm}\imag\int [dk] \frac{1}{(p_b-k_1)^2 k_1^2 (k_1-k_2)^2
(k_3^2-m_b^2) \left((k_2-k_3)^2-m_b^2\right)
\left((p_b-k_2)^2-m_b^2\right)}
\\
&=& \nonumber \frac{m_b^{-6\epsilon}}{2(4
\pi)^5}\left[\frac{1}{\epsilon ^2}+\frac{5}{\epsilon}-\frac{\pi ^2}{12}-4 \zeta (3)+15\right] +\mathcal{O}(\epsilon^1),
\end{eqnarray}
\begin{eqnarray}
&&\hspace{-1.8cm}\imag\int [dk] \frac{1}{(p_b-k_1)^2 k_1^2 (p_b-k_2)^2
k_2^2 \left((k_2-k_3)^2-m_b^2\right) \left((k_1-k_3)^2-m_b^2\right)}
\\
&=& \nonumber \frac{m_b^{-6\epsilon}}{2(4\pi)^5}\left[\frac{1}{2\epsilon^2}
+\frac{7}{2\epsilon}+\frac{33}{2}+\pi^2\left(-\frac{17}{24}-\frac{1}{\sqrt{5}}-\frac{4}{5}\log\frac{1+\sqrt{5}}{2}\right)+\frac{4\zeta(3)}{5}\right] +\mathcal{O}(\epsilon^1),
\end{eqnarray}
\begin{eqnarray}
&&\hspace{-1.6cm}\imag \int [dk] \frac{k_3^2}{(p_b-k_1)^2 k_1^2
(p_b-k_2)^2 k_2^2 \left((k_2-k_3)^2-m_b^2\right)
\left((k_1-k_3)^2-m_b^2\right)}
\\
&=& \nonumber \frac{m_b^{2-6\epsilon}}{2(4
\pi)^5}\left[\frac{13}{8\epsilon^2}+\frac{149}{16\epsilon}+\frac{1203}{32}-\pi^2\left(\frac{157}{96}+\frac{7}{4\sqrt{5}}+\frac{2}{5}\log\frac{1+\sqrt{5}}{2}\right)+\frac{2\zeta(3)}{5}\right] +\mathcal{O}(\epsilon^1).
\end{eqnarray}


\begin{thebibliography}{99}
\bibitem{hfag}
 \emph{Heavy Flavor Averaging Group (HFLAV)},
 \href{http://www.slac.stanford.edu/xorg/hfag/osc/summer_2017}{http://www.slac.stanford.edu/xorg/hfag/osc/summer\_2017.}

\bibitem{Abulencia:2006ze}
  A.~Abulencia {\it et al.} [CDF Collaboration],
  ``Observation of $B^0_s - \bar{B}^0_s$ Oscillations,''
  \href{http://dx.doi.org/10.1103/PhysRevLett.97.242003}
  {Phys.\ Rev.\ Lett.\  {\bf 97} (2006) 242003}
  \href{https://arxiv.org/abs/hep-ex/0609040}{[hep-ex/0609040].}

\bibitem{Aaij:2013mpa}
  R.~Aaij {\it et al.} [LHCb Collaboration],
  ``Precision measurement of the $B^{0}_{s}$-$\bar{B}^{0}_{s}$ oscillation frequency with the decay $B^{0}_{s}\rightarrow D^{-}_{s}\pi^{+}$,''
  \href{http://dx.doi.org/10.1088/1367-2630/15/5/053021}{New J.\ Phys.\  {\bf 15} (2013) 053021}
  \href{https://arxiv.org/abs/1304.4741}{[arXiv:1304.4741].}

\bibitem{lhcb}
  R.~Aaij {\it et al.} [LHCb Collaboration],
  ``Precision measurement of $CP$ violation in $B_s^0 \to J/\psi K^+K^-$ decays,''
  \href{http://dx.doi.org/10.1103/PhysRevLett.114.041801}{Phys.\ Rev.\ Lett.\  {\bf 114} (2015) no.4,  041801}
  \href{https://arxiv.org/abs/1411.3104}{[arXiv:1411.3104].}

 \bibitem{lhcb2}
   R.~Aaij {\it et al.} [LHCb Collaboration],
  ``First study of the CP -violating phase and decay-width difference
  in $B_s^0\to\psi(2S)\phi$ decays,''
   \href{http://dx.doi.org/10.1016/j.physletb.2016.09.028}{Phys.\ Lett.\ B {\bf 762} (2016) 253}
   \href{https://arxiv.org/abs/1608.04855}{[arXiv:1608.04855].}

\bibitem{Aad:2016tdj}
  G.~Aad {\it et al.} [ATLAS Collaboration],
  ``Measurement of the CP-violating phase $\phi_s$ and the $B^0_s$
  meson decay width difference with $B^0_s \to J/\psi\phi$ decays in ATLAS,''
   \href{http://dx.doi.org/10.1007/JHEP08(2016)147}{JHEP {\bf 1608} (2016) 147}
   \href{https://arxiv.org/abs/1601.03297}{[arXiv:1601.03297].}

\bibitem{Khachatryan:2015nza}
  V.~Khachatryan {\it et al.} [CMS Collaboration],
  ``Measurement of the CP-violating weak phase $\phi_s$ and the decay
  width difference $\Delta \Gamma_s$ using the B$_s^0 \to
  J/\psi\phi$(1020) decay channel in pp collisions at $\sqrt{s}=$ 8 TeV,''
  \href{http://dx.doi.org/10.1016/j.physletb.2016.03.046}{Phys.\ Lett.\ B {\bf 757} (2016) 97}
  \href{https://arxiv.org/abs/1507.07527}{[arXiv:1507.07527].}

\bibitem{Aaltonen:2012ie}
  T.~Aaltonen {\it et al.} [CDF Collaboration],
  ``Measurement of the Bottom-Strange Meson Mixing Phase in the Full CDF Data Set,''
  \href{http://dx.doi.org/10.1103/PhysRevLett.109.171802}{Phys.\ Rev.\ Lett.\  {\bf 109} (2012) 171802}
  \href{https://arxiv.org/abs/1208.2967}{[arXiv:1208.2967].}

\bibitem{Beneke:2003az}
  M.~Beneke, G.~Buchalla, A.~Lenz and U.~Nierste,
  ``CP asymmetry in flavor specific B decays beyond leading logarithms,''
  \href{http://dx.doi.org/10.1016/j.physletb.2003.09.089}{Phys.\ Lett.\ B {\bf 576} (2003) 173}
  \href{https://arxiv.org/abs/hep-ph/0307344}{[hep-ph/0307344].}

\bibitem{NiersteNLONB}
  A.~Lenz and U.~Nierste,
  ``Theoretical update of $B_s - \bar{B}_s$ mixing,''
  \href{http://dx.doi.org/10.1088/1126-6708/2007/06/072}{JHEP {\bf 0706} (2007) 072}
  \href{https://arxiv.org/abs/hep-ph/0612167}{[hep-ph/0612167].}

\bibitem{Lenz:2011ti} A.~Lenz and U.~Nierste, \emph{Proceedings of the
    6th International Workshop on the CKM Unitarity Triangle},
    Warwick, UK, 6-10 Sep 2010,
  ``Numerical Updates of Lifetimes and Mixing Parameters of B Mesons,''
  \href{https://arxiv.org/abs/1102.4274}{[arXiv:1102.4274].}

\bibitem{Nierste:2012qp}
  U.~Nierste,
  \emph{Proceedings of the 7th International Workshop on the CKM
    Unitarity Triangle}, Cincinnati, USA, 28 Sep - 2 Oct 2012,
  ``B Mixing in the Standard Model and Beyond,''
  \href{https://arxiv.org/abs/1212.5805}{[arXiv:1212.5805].}

\bibitem{hqe}
M.~A.~Shifman and M.~B.~Voloshin, in: \emph{Heavy Quarks}\ ed.\
V.~A.~Khoze and M.~A.~Shifman,
``Heavy Quarks,''
\href{http://dx.doi.org/10.1070/PU1983v026n05ABEH004398}{Sov.\ Phys.\ Usp.\  {\bf 26} (1983) 387.}

\bibitem{hqe2}
M.~A.~Shifman and M.~B.~Voloshin,
``Preasymptotic Effects In Inclusive Weak Decays Of Charmed Particles,''
Sov.\ J.\ Nucl.\ Phys.\  {\bf 41} (1985) 120
[Yad.\ Fiz.\  {\bf 41} (1985) 187];

\bibitem{hqe3}
M.~A.~Shifman and M.~B.~Voloshin,
``Hierarchy Of Lifetimes Of Charmed And Beautiful Hadrons,''
Sov.\ Phys.\ JETP {\bf 64} (1986) 698
[Zh.\ Eksp.\ Teor.\ Fiz.\  {\bf 91} (1986) 1180];

\bibitem{hqe4}
I.~I.~Bigi, N.~G.~Uraltsev and A.~I.~Vainshtein,
``Nonperturbative corrections to inclusive
beauty and charm decays: QCD versus phenomenological models,''
\href{http://dx.doi.org/10.1016/0370-2693(92)90908-M}{Phys.\ Lett.\ B {\bf 293} (1992) 430}
\href{https://arxiv.org/abs/hep-ph/9207214}{[hep-ph/9207214]}
\href{http://dx.doi.org/10.1016/0370-2693(92)91287-J}{[Erratum ibid. Phys.\ Lett.\ B {\bf 297} (1992) 477].}

\bibitem{HNSBsBsbar}
M.~Beneke, G.~Buchalla and I.~Dunietz,
  ``Width Difference in the $B_s-\bar{B}_s$ System,''
  \href{http://dx.doi.org/10.1103/PhysRevD.54.4419}{Phys.\ Rev.\ D {\bf 54} (1996) 4419}
   \href{https://arxiv.org/abs/hep-ph/9605259}{[hep-ph/9605259]}
   \href{http://dx.doi.org/10.1103/PhysRevD.83.119902}{[Erratum ibid. Phys.\ Rev.\ D {\bf 83} (2011) 119902].}

\bibitem{NiersteNLO}
 M.~Beneke, G.~Buchalla, C.~Greub, A.~Lenz and U.~Nierste,
  ``Next-to-leading order QCD corrections to the lifetime difference of $B_s$ mesons,''
  \href{http://dx.doi.org/10.1016/S0370-2693(99)00684-X}{Phys.\ Lett.\ B {\bf 459} (1999) 631}
  \href{https://arxiv.org/abs/hep-ph/9808385}{[hep-ph/9808385].}

\bibitem{Ciuchini:2003ww}
  M.~Ciuchini, E.~Franco, V.~Lubicz, F.~Mescia and C.~Tarantino,
  ``Lifetime differences and CP violation parameters of neutral B mesons at the next-to-leading order in QCD,''
  \href{http://dx.doi.org/10.1088/1126-6708/2003/08/031}{JHEP {\bf 0308} (2003) 031}
  \href{https://arxiv.org/abs/hep-ph/0308029}{[hep-ph/0308029].}

\bibitem{Bazavov:2016nty}
A.~Bazavov {\it et al.} [Fermilab Lattice and MILC Collaborations],
  ``$B^0_{s}$-mixing matrix elements from lattice QCD for the
  Standard Model and beyond,''
  \href{http://dx.doi.org/10.1103/PhysRevD.93.113016}{Phys.\ Rev.\ D {\bf 93} (2016) no.11, 113016}
  \href{https://arxiv.org/abs/1602.03560}{[arXiv:1602.03560].}

\bibitem{Badin:2007bv}
  A.~Badin, F.~Gabbiani and A.~A.~Petrov,
  ``Lifetime difference in $B_s$ mixing: Standard model and beyond,''
  \href{http://dx.doi.org/10.1016/j.physletb.2007.07.049}{Phys.\ Lett.\ B {\bf 653} (2007) 230}
  \href{https://arxiv.org/abs/0707.0294}{[arXiv:0707.0294].}

\bibitem{Buchalla:1995vs}
  G.~Buchalla, A.~J.~Buras and M.~E.~Lautenbacher,
  ``Weak decays beyond leading logarithms,''
  \href{http://dx.doi.org/10.1103/RevModPhys.68.1125}{Rev.\ Mod.\ Phys.\  {\bf 68}, 1125 (1996)}
  \href{https://arxiv.org/abs/hep-ph/9512380}{[hep-ph/9512380].}

 \bibitem{Gorbahn:2004my}
  M.~Gorbahn and U.~Haisch,
  ``Effective Hamiltonian for non-leptonic $|\Delta F| = 1$ decays at NNLO in QCD,''
  \href{http://dx.doi.org/10.1016/j.nuclphysb.2005.01.047}{Nucl.\ Phys.\ B {\bf 713} (2005) 291}
  \href{https://arxiv.org/abs/hep-ph/0411071}{[hep-ph/0411071].}

\bibitem{Gorbahn:2005sa}
  M.~Gorbahn, U.~Haisch and M.~Misiak,
  ``Three-loop mixing of dipole operators,''
   \href{http://dx.doi.org/10.1103/PhysRevLett.95.102004}{Phys.\ Rev.\ Lett.\  {\bf 95} (2005) 102004}
   \href{https://arxiv.org/abs/hep-ph/0504194}{[hep-ph/0504194].}

\bibitem{Brodsky:1982gc}
  S.~J.~Brodsky, G.~P.~Lepage and P.~B.~Mackenzie,
  ``On the Elimination of Scale Ambiguities in Perturbative Quantum Chromodynamics,''
  \href{http://dx.doi.org/10.1103/PhysRevD.28.228}{Phys.\ Rev.\ D {\bf 28}, 228 (1983).}

\bibitem{Beneke:1994qe}
  M.~Beneke and V.~M.~Braun,
  ``Naive nonabelianization and resummation of fermion bubble chains,''
  \href{http://dx.doi.org/10.1016/0370-2693(95)00184-M}{Phys.\ Lett.\ B {\bf 348}, 513 (1995)}
  \href{https://arxiv.org/abs/hep-ph/9411229}{[hep-ph/9411229].}

\bibitem{Asatrian:2010rq}
  H.~M.~Asatrian, T.~Ewerth, A.~Ferroglia, C.~Greub and G.~Ossola,
  ``Complete $(O_7,O_8)$ contribution to $\bar{B}\to X_s \gamma$ at order $\alpha_s^2$,''
   \href{http://dx.doi.org/10.1103/PhysRevD.82.074006}{Phys.\ Rev.\ D {\bf 82}, 074006 (2010)}
   \href{https://arxiv.org/abs/1005.5587}{[arXiv:1005.5587].}

\bibitem{Beneke:2002rj}
  M.~Beneke, G.~Buchalla, C.~Greub, A.~Lenz and U.~Nierste,
  ``The $B^+ - B^0_d$ lifetime difference beyond leading logarithms,''
  \href{http://dx.doi.org/10.1016/S0550-3213(02)00561-8}{Nucl.\ Phys.\ B {\bf 639} (2002) 389}
  \href{https://arxiv.org/abs/hep-ph/0202106}{[hep-ph/0202106].}

\bibitem{Buras:2006gb}
  A.~J.~Buras, M.~Gorbahn, U.~Haisch and U.~Nierste,
  ``Charm quark contribution to $K^+ \to \pi^+ \nu \bar{\nu}$ at next-to-next-to-leading order,''
  \href{http://dx.doi.org/10.1088/1126-6708/2006/11/002}{JHEP {\bf 0611}, 002 (2006)}
  \href{http://dx.doi.org/10.1007/JHEP11(2012)167}{[JHEP {\bf 1211}, 167 (2012)]}
  \href{https://arxiv.org/abs/hep-ph/0603079}{[hep-ph/0603079].}

  \bibitem{NiersteEvan}
S.~Herrlich and U.~ Nierste, ``Evanescent operators, scheme dependences and double insertions,'' \href{http://dx.doi.org/10.1016/0550-3213(95)00474-7}{Nucl.\ Phys.\ {\bf B 455} 39 (1995)}
\href{https://arxiv.org/abs/hep-ph/9412375}{[hep-ph/9412375].}

\bibitem{Chetyrkin:1997fm}
  K.~G.~Chetyrkin, M.~Misiak and M.~Munz,
  ``Beta functions and anomalous dimensions up to three loops,''
  \href{http://dx.doi.org/10.1016/S0550-3213(98)00122-9}{Nucl.\ Phys.\ B {\bf 518}, 473 (1998)}
  \href{https://arxiv.org/abs/hep-ph/9711266}{[hep-ph/9711266].}

\bibitem{Bieri:2003ue}
  K.~Bieri, C.~Greub and M.~Steinhauser,
  ``Fermionic NNLL corrections to $b \to s \gamma$,''
  \href{http://dx.doi.org/10.1103/PhysRevD.67.114019}{Phys.\ Rev.\ D {\bf 67}, 114019 (2003)}
  \href{https://arxiv.org/abs/hep-ph/0302051}{[hep-ph/0302051].}

\bibitem{bw}
 A.~J.~Buras and P.~H.~Weisz,
  ``QCD Nonleading Corrections to Weak Decays in Dimensional
  Regularization and 't Hooft-Veltman Schemes,''
  \href{http://dx.doi.org/10.1016/0550-3213(90)90223-Z}{Nucl.\ Phys.\ B {\bf 333} (1990) 66.}

\bibitem{Gray:1990yh}
  N.~Gray, D.~J.~Broadhurst, W.~Grafe and K.~Schilcher,
  ``Three-loop relation of quark $\overline{MS}$ and pole masses,''
  \href{http://dx.doi.org/10.1007/BF01614703}{Z.\ Phys.\ C {\bf 48} (1990) 673.}

\bibitem{Chetyrkin:1999qi}
  K.~G.~Chetyrkin and M.~Steinhauser,
  ``The Relation between the $\overline{MS}$ and the on-shell quark mass at
  order $\alpha_s^3$,''
  \href{http://dx.doi.org/10.1016/S0550-3213(99)00784-1}{Nucl.\ Phys.\ B {\bf 573} (2000) 617}
  \href{https://arxiv.org/abs/hep-ph/9911434}{[hep-ph/9911434].}

\bibitem{Asatrian:2006rq}
  H.~M.~Asatrian, T.~Ewerth, H.~Gabrielyan and C.~Greub,
  ``Charm quark mass dependence of the electromagnetic dipole operator contribution to $\bar B \to X_s \gamma$ at $O(\alpha_s^2)$,''
  \href{http://dx.doi.org/10.1016/j.physletb.2007.02.027}{Phys.\ Lett.\ B {\bf 647}, 173 (2007)}
  \href{https://arxiv.org/abs/hep-ph/0611123}{[hep-ph/0611123].}

\bibitem{Agashe:2014kda}
C.~Patrignani {\it et al.} [Particle Data Group], ``Review of Particle Physics,''
\href{http://dx.doi.org/10.1088/1674-1137/40/10/100001}{Chin.\ Phys.\ C {\bf 40}, no. 10, 100001 (2016).}

\bibitem{Charles:2004jd}
  J.~Charles {\it et al.}  [CKMfitter Group Collaboration],
  ``CP violation and the CKM matrix: Assessing the impact of the asymmetric $B$ factories,''
  \href{http://dx.doi.org/10.1140/epjc/s2005-02169-1}{Eur.\ Phys.\ J.\ C {\bf 41}, 1 (2005)}
  \href{https://arxiv.org/abs/hep-ph/0406184}{[hep-ph/0406184].}
  %
  We use updated numbers from  \href{http://ckmfitter.in2p3.fr}{http://ckmfitter.in2p3.fr.}

\bibitem{Kuhn:2007vp}
  J.~H.~Kuhn, M.~Steinhauser and C.~Sturm,
  ``Heavy Quark Masses from Sum Rules in Four-Loop Approximation,''
   \href{http://dx.doi.org/10.1016/j.nuclphysb.2007.04.036}{Nucl.\ Phys.\ B {\bf 778} (2007) 192}
  \href{https://arxiv.org/abs/hep-ph/0702103}{[hep-ph/0702103].}

\bibitem{Allison:2008xk}
  I.~Allison {\it et al.} [HPQCD Collaboration],
  ``High-Precision Charm-Quark Mass from Current-Current Correlators in Lattice and Continuum QCD,''
   \href{http://dx.doi.org/10.1103/PhysRevD.78.054513}{Phys.\ Rev.\ D {\bf 78} (2008) 054513}
   \href{https://arxiv.org/abs/0805.2999}{[arXiv:0805.2999].}

\bibitem{Aoki:2016frl}
  S.~Aoki {\it et al.},
  ``Review of lattice results concerning low-energy particle physics,''
   \href{http://dx.doi.org/10.1140/epjc/s10052-016-4509-7}{Eur.\ Phys.\ J.\ C {\bf 77} (2017) no.2,  112}  \href{https://arxiv.org/abs/1607.00299}{[arXiv:1607.00299].}

\bibitem{Huber:2005ig}
  T.~Huber, E.~Lunghi, M.~Misiak and D.~Wyler,
  ``Electromagnetic logarithms in $\bar{B}\to X_s l^+ l^-$,''
  \href{http://dx.doi.org/10.1016/j.nuclphysb.2006.01.037}{Nucl.\ Phys.\ B {\bf 740}, 105 (2006)}
  \href{https://arxiv.org/abs/hep-ph/0512066}{[hep-ph/0512066].}

\bibitem{Smirnov:2008iw}
  A.~V.~Smirnov,
  ``Algorithm FIRE -- Feynman Integral REduction,''
  \href{http://dx.doi.org/10.1088/1126-6708/2008/10/107}{JHEP {\bf 0810}, 107 (2008)}
  \href{https://arxiv.org/abs/0807.3243}{[arXiv:0807.3243].}

\bibitem{Asatrian:2012tp}
  H.~M.~Asatrian, A.~Hovhannisyan and A.~Yeghiazaryan,
  ``The phase space analysis for three and four massive particles in final states,''
  \href{http://dx.doi.org/10.1103/PhysRevD.86.114023}{Phys.\ Rev.\ D {\bf 86}  (2012) 114023}
  \href{https://arxiv.org/abs/1210.7939}{[arXiv:1210.7939].}

\bibitem{Boos:1990rg}
  E.~E.~Boos and A.~I.~Davydychev,
  ``A Method of evaluating massive Feynman integrals,''
   \href{http://dx.doi.org/10.1007/BF01016805}{Theor.\ Math.\ Phys.\  {\bf 89}, 1052 (1991).}

\bibitem{Borowka:2015mxa}
  S.~Borowka, G.~Heinrich, S.~P.~Jones, M.~Kerner, J.~Schlenk and T.~Zirke,
  ``SecDec-3.0: numerical evaluation of multi-scale integrals beyond one loop,''
  \href{http://dx.doi.org/10.1016/j.cpc.2015.05.022}{Comput. Phys. Commun.  {\bf 196}, 470 (2015)}
  \href{https://arxiv.org/abs/1502.06595}{[arXiv:1502.06595].}





\end{thebibliography}
\end{document}